\documentclass[preprint,12pt,authoryear]{elsarticle}

\usepackage{amsmath,amssymb}
\usepackage{float}
\usepackage{subcaption}
\usepackage{xcolor}
\usepackage{graphicx}
\usepackage{multirow}
\usepackage{url}

\journal{Renewable Energy}

\begin{document}

\begin{frontmatter}



\title{Multi-Objective Multidisciplinary Optimization of Wave Energy Converter Array Layout and Controls}

\author{\textbf{Kapil Khanal$^{a}$, Nate DeGoede$^{b}$, Olivia Vitale$^{b}$, Maha N. Haji$^{a,b}$}}


\affiliation[a]{organization={Systems Engineering, Cornell University},
            city={Ithaca},
            state={NY},
            country={USA}}

\affiliation[b]{organization={Mechanical and Aerospace Engineering, Cornell University},
            city={Ithaca},
            state={NY},
            country={USA}}

\begin{abstract}
This study utilizes multidisciplinary design optimization (MDO) to design an array of heaving wave energy converters (WECs) for grid-scale energy production with decision variables and parameters chosen from the coupled disciplines of geometry, hydrodynamics, layout, motor-actuated reactive controls (with a force maximum constraint) and economics. We vary a WEC's dimensions, array layout, and control gain to minimize two objectives: the levelized cost of energy (LCOE) and the maximum separation distance. This multi-objective optimization approach results in a set of optimal design configurations that stakeholders can choose from for their specific application and needs. The framework yields a range of optimal (minimum) LCOE values from 0.21 to 0.23 \$/kWh and a separation distance ranging from 97 to 62 meters. The WEC radius of 4m is found to be optimal, and the q-factor for optimal designs are greater than 1 up to 1.06. The optimal configuration for 4 heaving WECs is found to be a rhombus-like layout. Additionally, a post-optimality global sensitivity analysis is performed to analyze the interaction of several parameters. The uncertainty (variance) of the LCOE is quantified in terms of the uncertainty of the input parameters and interpreted to aid in decision making. Wave heading, wave frequency, WEC lifetime, amplitude and interest rate accounts for most of the variance in the optimal LCOE configuration. Different designs in the Pareto set may be appealing for different decision makers based on their trade-off analysis. To that end, an ad-hoc regression model is developed for trade-off analysis between objectives and some design heuristics are provided.
\end{abstract}

\begin{graphicalabstract}
\includegraphics[width = 1.2\textwidth]{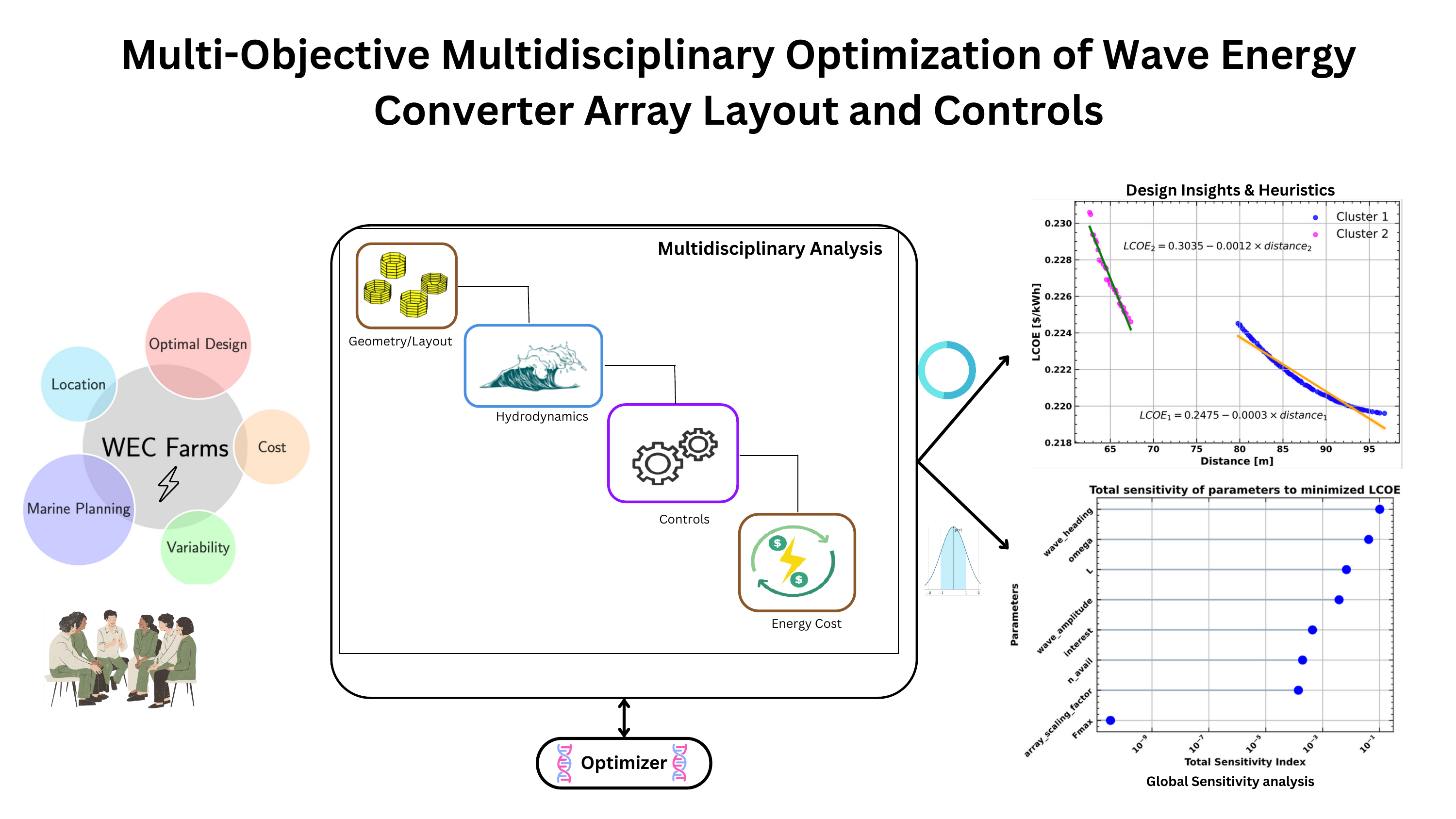}
\end{graphicalabstract}

\begin{highlights}
\item Optimal designs differ by 37\% in ocean space used and 5\% in LCOE.
\item The q-factor for optimal designs range between  1 and 1.06
\item Optimal configurations for 4 heaving WECs is rhombus-like shape
\item Regression model interprets the trade-off between objectives for optimal designs
\item Sensitivity analysis suggests mild sensitivity to wave environment and interest rate

\end{highlights}


\end{frontmatter}


\section{Introduction}
Wave energy converters (WECs) convert the oscillatory motion of ocean waves into usable mechanical or electrical energy \citep{Budal1977}. Multiple WECs are required at a site for grid scale integration, motivating the study of WEC arrays, or farms. Wave farms have the potential to supply 34\% of the US electric supply needs \citep{Kilcher2021}. However, implementing these offshore systems is expensive and challenging. Aside from the bureaucratic and permitting hurdles that exist for developing WEC farms outlined in the study by \cite{pacwave}, many technical challenges still face the industry. Babarit et al detailed the ``park effect" of WEC arrays, where the waves radiated and diffracted from each body affect the motion of all interacting bodies, thereby affecting total farm power output \citep{BABARIT201368}. Designing a farm configuration that increases power output requires detailed numerical models that account for these interactions. Computational expenses for the numerical solver scales cubically with the number of WEC bodies, making numerical hydrodynamic modeling difficult for large arrays \citep{newman1985algorithms}. Several modeling methods exist for various types of array investigations (\cite{folley2012review}, \cite{YANG2022112668}), but there is no universally applicable model. Numerical solutions along with a series of design studies for the trade-off analysis between competing objectives are necessary for WEC farms. 

When optimizing WEC arrays, the number of WECs, separation distance, array layout, and wave environment are shown to be important factors for WEC farms (\cite{DEANDRES201432}) and thus must be included in early design studies. \cite{BOZZI2017378} simulated WEC array configurations in real-wave conditions off the Italian coastline. They highlighted the importance of dominant wave heading on determining the optimal configuration; however, the optimal designs found were based on brute-force optimization for maximizing power using only four configurations. \cite{LYU2019106543} used a genetic algorithm to optimize 3, 5, and 7-body arrays for configuration and body dimension, finding up to a 39\% increase in power production when optimizing both design variables. They also emphasize that the control method influences the optimal array. However, their optimal designs include WECs of the same architecture with slightly different radii and drafts, making the design impractical from an economics of scale standpoint. Most models in literature optimize solely for power production \citep{BABARIT201244},\citep{goteman2015geometries},\citep{ABDULKADIR2023113818}, ignoring other maritime priorities. 

Marine spatial planning (MSP) is an important factor when attempting marine developments. Managing space for all maritime markets (fishing, renewable energy, marine conservation, shipping routes, etc.) is required when creating offshore developments (\cite{EHLER2021104134}). It is also imperative to keep WECs within an optimal siting location because of ecological considerations such as disturbances, noise and many ecosystem elements (\cite{framework_ecology}).  Ensuring the farm can be practically sited and coexist with other maritime markets is essential for their successful real-world implementation. The siting location and ocean space required is influenced by the distances between the WECs in a farm. Many studies find a distance as a function of WEC diameter where bodies should be placed to avoid interactions (\cite{BOZZI2017378}, \cite{BABARIT201368}, \cite{SINGH201412}). However, these distances are large (greater than 10 times the WEC diameter) and may be impractical for arrays from an operation, maintenance and mooring infrastructure development point of view. Thus, exploiting array effects in compact configurations is crucial for bringing WEC farms to fruition.

This optimization study analyzes economic viability, ocean space used, and power production for a 4-body WEC farm in regular waves. Economics are reported through the levelized cost of energy (LCOE), which measures the cost of energy production over the lifetime of the system and provides a benchmark for cost comparison (\cite{nahvi_2018}). The ocean space used is characterized by the farm length; the maximum distance between any two WECs. LCOE and farm length are both minimized in the optimization. For additional analysis (not directly optimized), the power production for the optimal designs is also reported in terms of q-factor. Q-factor is a metric used to compare the total power production of an $n$-body array to the power production of $n$ isolated WECs. These objectives depend on WEC geometry and layout, hydrodynamics, controls, and economics. All these disciplines rely on each other, making this a high dimensional multidisciplinary system design problem. Individual WECs have been already designed in multidisciplinary way \citep{mccabe_multidisciplinary_2022}. This study introduces the application of the MDO technique to the array layout optimization. Such a formulation is considered to provide better designs than conventional sequential design optimization \citep{mdobook}.

In addition to the optimization, a sensitivity study to explore the interaction of the parameters between different technical disciplines is of vital importance. Understanding these relationships is not just a mathematical problem but also will help accelerate adoption and permitting of engineered systems. A study conducted at the PacWave wave energy test facility found that addressing uncertainty regarding wave technologies is one of several areas of utmost importance for the most stakeholders \citep{pacwave}. Similarly, a report by Sandia national lab points out that the WEC farms readiness has been too focused on commercial readiness and not enough on the stakeholder requirements and economic viability required for market entry \citep{osti_1365534}. They document that among many requirements for successful wave energy farms, it is essential that they are economically competitive and address social considerations. In this paper, a global sensitivity analysis (GSA) is conducted for the sensitivity calculation to the optimal result from optimization study with many objectives, coming from different stakeholders. This technique captures the global behaviour of the models and has been applied to many engineering systems such as wind farms \citep{wind_global}.

Uncertain ocean parameters affect the power production of a WEC, and should therefore be quantified as they provide valuable information on location data and the expected variability of the power production. Metrics like LCOE are sensitive to environmental, dynamic and economic parameters. Various authors have looked into the local variation of these metrics to quantify their performance. \cite{Sinha2016} investigated the influence of incident angle on the power absorption and power consistency of different fixed optimally tuned array configurations. They found that for concentric and circular arrays, the power absorption profile remains the same while for linear and grid type arrays, the power absorption increases with increase in incident from 0 to 90 degrees suggesting sensitivity of incidence angle on array configurations. \cite{sharp2018wave} performed a minimum separation study and quantified the significance of the the separating distance on array performance metrics. Different restrictions on the distance resulted in different layouts. This could be partly attributed to simpler model \citep{mcnatt2015novel} they used for the hydrodynamic calculations. Using a more accurate hydrodynamic model may provide more insight into its interactions with other disciplines and the design space becomes much 'richer' for the power calculations. For example, interaction between wave environment and economics parameters are lacking in most of these studies. Similarly, rather than solely performing limited local parameter studies on a few common configurations, it is essential to integrate sensitivity analysis with the layout optimization studies within a complete multidisciplinary model of the WEC farms, as considered in this paper. 

A global sensitivity analysis based on Sobol indices and variance decomposition  \citep{saltelli1993sensitivity} is used to calculate the sensitivity of most important parameters for the recommended Pareto optimal design. A variance-based uncertainty quantification is used because it is model-independent and can capture global interactions between input parameters. This also means that this method can be coupled with other software frameworks for hydrodynamic and optimization analysis.
Sensitivity analysis is performed for the recommended Pareto optimal design, partly due to complexity for the disciplinary analysis such as the numerical simulation. The curse of dimensionality prevents the use of this method within an optimization. The approach we take enhances the understanding of the Pareto designs by providing the sensitivities. 

In Section 2, we discuss the simulation model, and optimization formulation, and Section 3 describes the validation of the hydrodynamic model. Section 4 discusses the results of the optimization, and Section 5 discusses the post-optimality sensitivity analysis, and Section 6 discusses the future work.

\section{Methodology}
In this section, we discuss the problem formulation and introduce a design structure matrix of the multidisciplinary model consisting of several modules. Each individual module and its coupling with other disciplines is discussed. 

\subsection{Problem Formulations}

In standard MDO problem formulations, each discipline is modeled as a separate module. The overall problem includes one or more objectives and constraints with multiple design variables and parameters. The inter-dependency between disciplinary modules are shown via coupling variables which can result in both feedforward and feedback loops. A multidisciplinary feasibility (MDF) is then obtained for the coupled disciplines \citep{cramer1994problem}. The formal MDO definition of the problem discussed in this paper is given by


\begin{align*} 
\text{Minimize} \quad & \begin{bmatrix} LCOE(\mathbf{x},\mathbf{p}; \hat{u}^) \\ SPACE_{\text{max}}(\mathbf{x},\mathbf{p}; \hat{u}^) \end{bmatrix}  \\
\text{by varying} \quad & \mathbf{x}\\
\text{subject to } \quad &  C_0 = 5r - \sqrt{(x_i - x_j)^2 + (y_i - y_j)^2}     \leq 0 \\
\text{while solving} \quad & R_i(\mathbf{x},\mathbf{p};\hat{u}) = 0\\
\text{for } \quad & \hat{u}
\label{eq:problem}
\end{align*}

 where $LCOE \textrm{ and } SPACE_\text{max} $ are the two objectives to be minimized, \textbf{x} is the vector of design variables. $r$ is the WEC radius (discretized in the increment of 0.5m), ($x_i$,$y_i$) and ($x_j$,$y_j$) are the WEC coordinates of which first WEC is always fixed at $(0,0)$. The vector of the coupled variables is $\hat{u}$. $R_i$ refers to the governing equations in any of the disciplines. Inequality constraint ($C_0$) ensures the WECs remain a distance of at least five times the radius apart. This value provides a large enough distance to prevent collisions between WECs, but also close enough that interactions between bodies will still affect the system in complex ways \citep{SINGH201412}. 
 
\begin{figure}[H]
    \centering
\includegraphics[width = 1.0\linewidth]{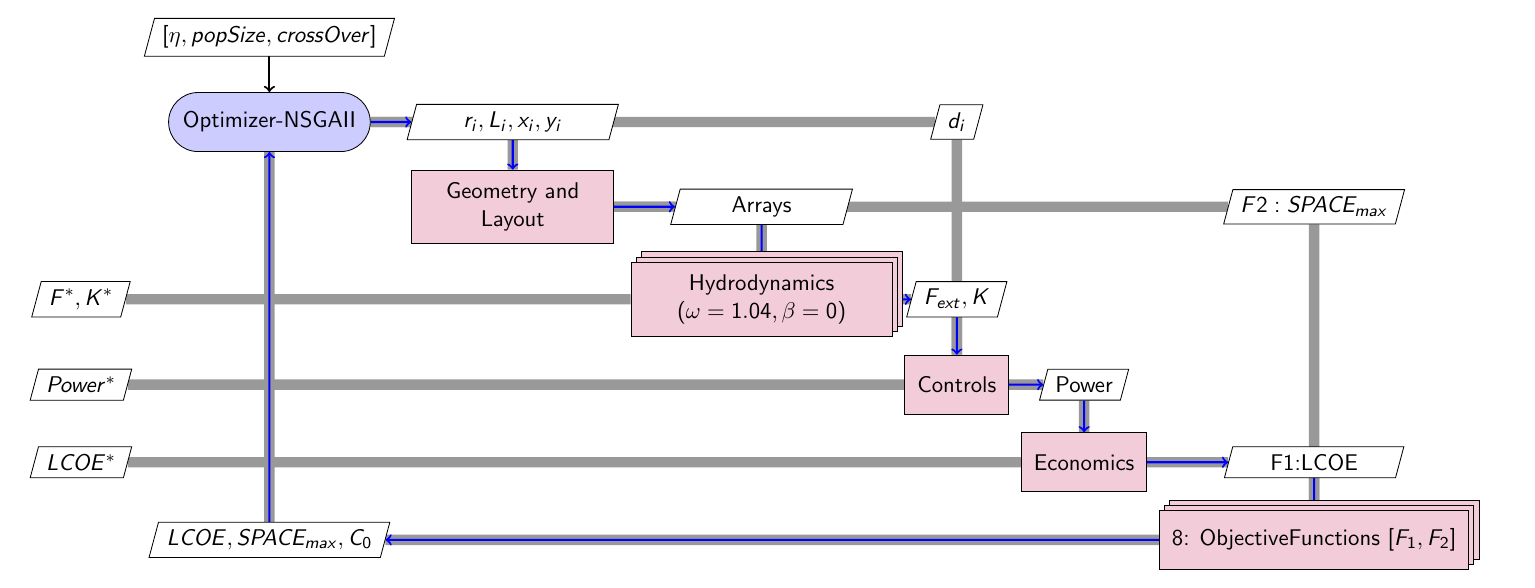}
    \caption{xDSM diagram of the multidisciplinary model of WEC farms. This architecture couples the optimizer with the multidisciplinary analysis of all the relevant modules. The diagonals are the disciplines and off-diagonals are the coupling variables. The input design variables are shown passing from the top and output of each discipline is shown on the right. The optimal value for each discipline is shown with an asterisk(*) of the left after the optimizer converges. The grey line shows the data I/O and blue line shows the sequence of analysis. }
    \label{fig:xdsm}
\end{figure}

\begin{figure}[ht!]
    \centering
    \includegraphics[width = 0.6\linewidth]{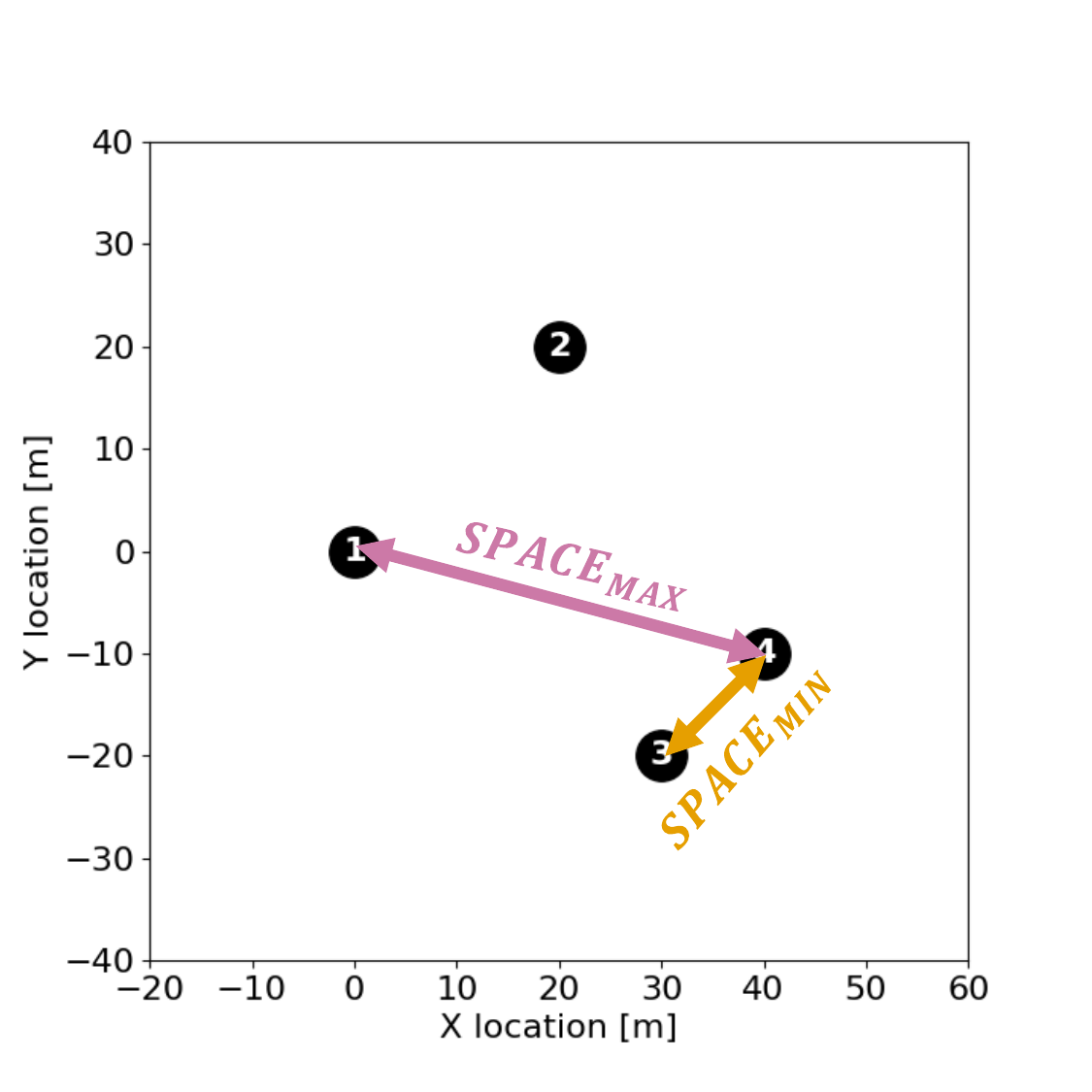}
    \caption{Plot showing the minimum space constraint and the SPACE$_{\text{max}}$ objective.}
    \label{fig:spacelabels}
\end{figure}

 A visualization of the space objective and constraint $C_0$ is show in Fig. \ref{fig:spacelabels} and Table \ref{table:p_x_g_j} details the parameters, design variables, constraints, and objectives. The xDSM \citep{xdsm} diagram (Fig. \ref{fig:xdsm}) depicts the simulation models, disciplines used, and their interactions. All coupling variables are either explicitly or implicitly the function of the design vectors. The disciplines are ordered to ensure that no feedback loops exist in the xDSM diagram. 

\begin{figure}[H]
    \centering
    \includegraphics[width = 0.5\linewidth]{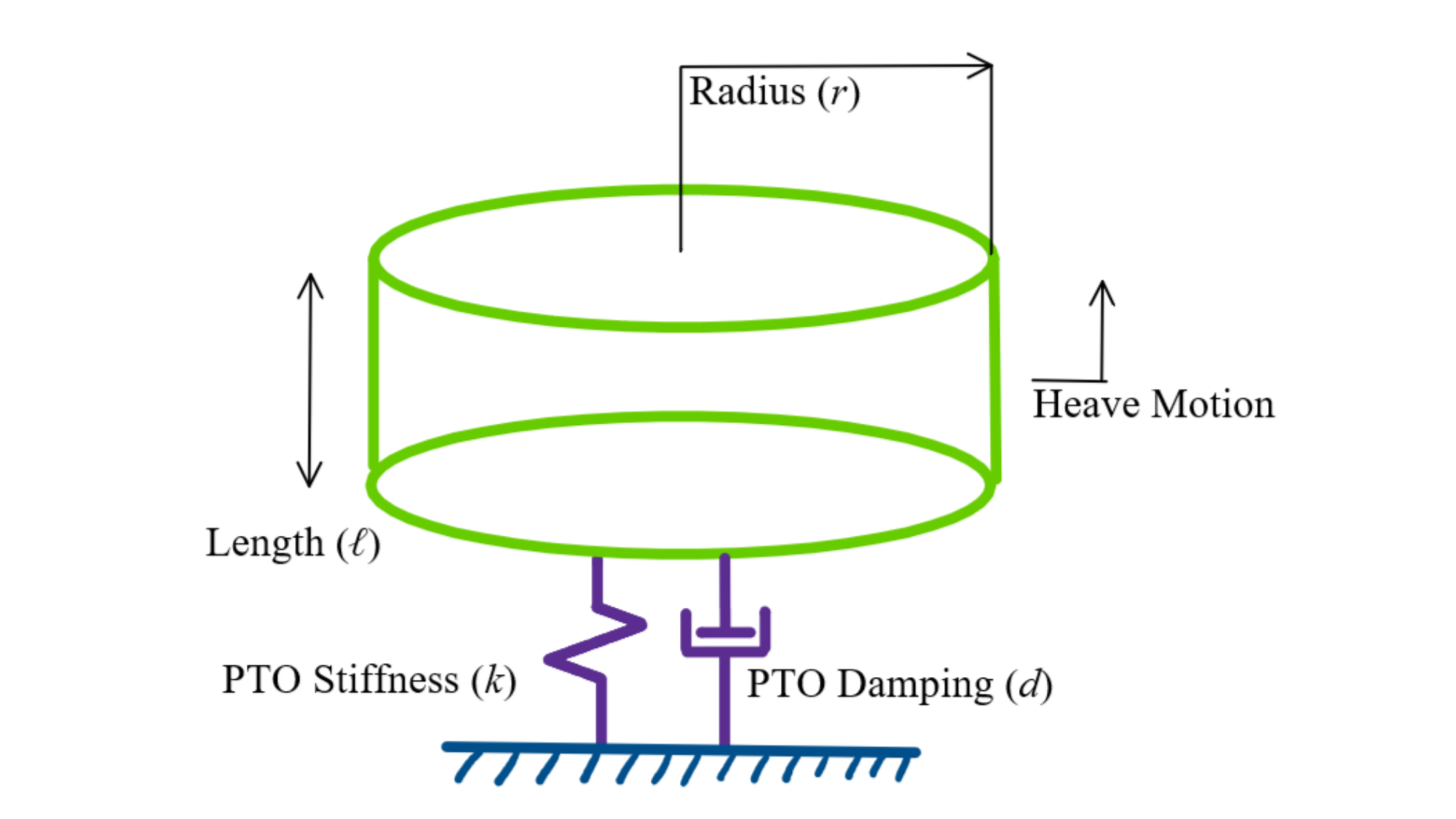}
    \caption{Schematic of the WEC geometry and control design variables.}
    \label{fig:wec}
\end{figure}

The input design vector is
\begin{equation}
    \mathbf{x} = [2r,\ell/r,log_{10}(d_1),x_2,y_2,log_{10}(d_2),x_3,y_3,log_{10}(d_3), ..., x_n,y_n,log_{10}(d_n)]^{T} 
    \label{eq:x}
\end{equation}

 where $\ell$ is the length of the WEC, $d$ is the damping of the power take-off system (PTO), and subscript $n$ represents the total number of WECs. Fig. \ref{fig:wec} shows the WEC geometry and control design variables. Also note that the design variable $2r$, representing diameter, is limited to integer values.

\begin{table}[ht!]
\caption{Main Table outlining the parameters, design variables, constraint, and objectives in this MDO formulation.}
\tiny
\centering
\begin{tabular}{|p{2cm}|p{3.3cm}|p{1.5cm}|p{4cm}|}
\hline
{\color[HTML]{000000} } & {\color[HTML]{000000} } & {\color[HTML]{000000} } & {\color[HTML]{000000} }  \\
\multirow{-2}{*}{{\color[HTML]{000000} \textbf{}}} & \multirow{-2}{*}{{\color[HTML]{000000} \textbf{Name}}} & \multirow{-2}{*}{{\color[HTML]{000000} \textbf{Symbol}}} & \multirow{-2}{*}{{\color[HTML]{000000} \textbf{Notes}}} \\ 
\hline
{\color[HTML]{000000} }  & {\color[HTML]{000000} Wave Frequency} & {\color[HTML]{000000} $\omega$} & {\color[HTML]{000000} 1.047 rad/s used in this study}   \\
{\color[HTML]{000000} }  & {\color[HTML]{000000} Wave Amplitude} & {\color[HTML]{000000} $A$} & {\color[HTML]{000000} 1 m used in this study} \\
{\color[HTML]{000000} }  & {\color[HTML]{000000} } & {\color[HTML]{000000} } & {\color[HTML]{000000} measured from +x axis} \\
{\color[HTML]{000000} }  & \multirow{-2}{*}{{\color[HTML]{000000} Wave Direction}} & \multirow{-2}{*}{{\color[HTML]{000000} $\beta$}} & 0 rad used in this study \\
{\color[HTML]{000000} }  & {\color[HTML]{000000} Interest Rate} & {\color[HTML]{000000} $i$} & {\color[HTML]{000000} 7\%}   \\
{\color[HTML]{000000} }  \multirow{-1}{*}{{\color[HTML]{000000} \textbf{Parameters}}} & {\color[HTML]{000000} Wave Resource Availability} & {\color[HTML]{000000} $\eta_{avail}$} & {\color[HTML]{000000} 95\%}   \\
{\color[HTML]{000000} }  & {\color[HTML]{000000} WEC Life Span} & {\color[HTML]{000000} $L$} & {\color[HTML]{000000} 25 years}   \\
{\color[HTML]{000000} }  & {\color[HTML]{000000} Array Scaling Factor} & {\color[HTML]{000000} $S$} & {\color[HTML]{000000} 0.65}   \\
{\color[HTML]{000000} }  & {\color[HTML]{000000} PTO Force Maximum} & {\color[HTML]{000000} $F_{max}$} & {\color[HTML]{000000} 2.6$\times 10^5$ N (optimum from \cite{penasanchez})}   \\
{\color[HTML]{000000} }  & {\color[HTML]{000000} Median Capital Cost} & {\color[HTML]{000000} $CAPEX_{med}$} & {\color[HTML]{000000} \$9000/kW}   \\

\hline
{\color[HTML]{000000} } & {\color[HTML]{000000} WEC Radius} & {\color[HTML]{000000} $r$} & {\color[HTML]{000000} ranges from 2 to 10 m} \\
{\color[HTML]{000000} } & WEC Length & $\ell$ & ranges from $0.1r$ to $2r$ \\
{\color[HTML]{000000} } &  &  & ranges from -500 m to 500 m \\
{\color[HTML]{000000} } & \multirow{-2}{*}{X Location} & \multirow{-2}{*}{$x_i$} & $x_1$ is fixed at 0 m \\
{\color[HTML]{000000} } &  &  & ranges from -500 m to 500 m \\
{\color[HTML]{000000} } & \multirow{-2}{*}{Y Location} & \multirow{-2}{*}{$y_i$} & $y_1$ is fixed at 0 \\
\multirow{-7}{*}{{\color[HTML]{000000} \textbf{Design Variables}}} & PTO Damping & $d_i$ & ranges from 1 to 10,000,000 Ns/m \\ 
\hline
{\textbf{Constraints}}& Minimum WEC Spacing & $SPACE_{min}$ & must be larger than $5r$ \\ 
\hline
& Levelized Cost of Energy & $LCOE$ & \$/kWh \\
\multirow{-2}{*}{\textbf{Objectives}} & Maximum WEC Spacing & $SPACE_{max}$ & m       \\
\hline
\end{tabular}
\label{table:p_x_g_j}
\end{table}

The parameters for this problem are

\begin{equation}
    \mathbf{p} = [\omega,A,\beta,i,\eta_{avail},L,S,F_{max},CAPEX_{med}]
\end{equation}

where $\omega$ is the incoming wave frequency, $A$ is the wave amplitude, $\beta$ is the wave direction, $i$ is the interest rate, $\eta_{avail}$ is the wave availability coefficient, $L$ is the anticipated WEC lifetime, $S$ is the array scaling factor, $F_{max}$ is the maximum force constraint on the PTO, and $CAPEX_{med}$ is the median capital cost as detailed in Section \ref{sec:econ}.

The following sections discuss each module in detail, including fundamental principles and equations as well as the optimization details, the sensitivity calculations and the discussion of the results.

\subsubsection{Optimization} \label{opt}
This study optimizes for two objectives using a multi-objective genetic algorithm (MOGA). A Pareto set is sought after to find the trade-off between the objectives. A non-dominated sorting genetic algorithm (NSGA-II) adapted for mixed variable types is used as an optimizer. NSGA-II is an evolutionary algorithm with elitist strategy, a parameter-less approach and efficient constraint-handling method (\cite{NSGAII}). We use a population size of 120  as well as the default hyperparameters utilized by the MixedVariableGA solver in the python library pymoo \citep{pymoo}. Robust termination criteria in the objective function is used for the convergence. The tolerance level of 0.005 change in both objectives is required before the algorithm terminates. 

\subsection{Geometry and Layout}
A heaving point absorber WEC is considered in this paper, and cost is directly related to its shape and size. Usually, the radius and draft  are optimized for a wave environment. In addition to the optimal sizing of WECs, their layout is also important because how the WECs are situated will affect the wave field significantly. Such a design will also have to satisfy the competing requirement to minimize the ocean space used. Optimal dimensions for different layout types, such as a regular or staggered grid, are explored in literature \citep{YANG2022112668}. The fixing of the topology of the layout and only optimizing its dimension provides some computational benefits. \cite{goteman2015geometries} analyzed over 1000 interacting point absorbers using an approximate analytic method for single body diffraction. They found that among many different layouts (wedges, rectangle, circles, random) the absorption power remains comparable, but the power variance differs significantly between configurations. This relation between power and layout could be vastly different if we were to include, for instance layout infrastructure costs, radiated waves and higher fidelity hydrodynamics analysis.

From the MDA perspective, the geometry module generates the WEC bodies on different locations as given by the optimizer. A heaving point absorber WEC is modeled as a simple cylinder (see Fig. \ref{fig:wec}) designed to float about its center of mass. The initial array layout and WEC positions are randomized to reduce influence on the heuristic optimization algorithm and find the most optimal conditions. 
 
 Our initial investigations show that WEC size and optimal PTO damping is dependent on array layout. We used a single objective heuristic optimization (with genetic algorithm) for LCOE finding the optimal geometry and control for WECs in two different layout types (grid and line, shown in Appendix Fig. \ref{fig:layouts}). The resulting optimal WEC geometry and dampings are shown in Table \ref{table:geom_damps_soo}.
 
\begin{table}[]
\centering
\caption{Optimal geometry and PTO dampings for the different geometries shown in Fig. \ref{fig:layouts}.}
\begin{tabular}{llllllll}
Layout & Spacing {[}m{]} & $r$ {[}m{]} & $\ell$ {[}m{]} & $d_1$ {[}$\frac{kNs}{m}${]} & $d_2$ {[}$\frac{kNs}{m}${]} & $d_3$ {[}$\frac{kNs}{m}${]} & $d_4$ {[}$\frac{kNs}{m}${]} \\ \hline
Grid & 50 & 3 & 0.300 & 129 & 129 & 129 & 130 \\
Grid & 70 & 4 & 0.400 & 378 & 382 & 378 & 378 \\
Line & 30 & 4 & 0.400 & 414 & 446 & 451 & 421 \\
Line & 70 & 4 & 0.400 & 367 & 401 & 406 & 398 
\end{tabular}
\label{table:geom_damps_soo}
\end{table}

\subsection{Hydrodynamics}


Wave-structure interactions are commonly characterized using linear potential flow theory. The theory assumes the fluid velocity field $\vec{v}$ is the gradient of some complex potential ($\phi$), where $\vec{v}=\nabla \phi$ and $\phi$ satisfies the Laplace equation, $\nabla^2 \phi = 0$. It also assumes inviscid, irrotational, and incompressible flow. $\phi(x,y,z)$ is complex velocity potential represented in $ \Phi = \mathbb{Re}(\phi e^{-j\omega t})$

The boundary conditions for the associated boundary value problem (BVP) are  

\begin{equation}\label{BCs}
\begin{aligned}
\left. \frac{\partial \phi}{\partial z} \right|_{z=-h} &= 0 \\ 
\\
\left. \frac{\partial \phi}{\partial z} - \frac{\omega^2}{g} \phi \right |_{z=0} & = 0 \\
\\
\left. \frac{\partial \phi}{\partial \vec{n}} \right |_\Gamma &= \vec{u} \cdot \vec{n} \\
\\
\lim_{R \to \infty} \left( \sqrt{kR} \frac{\partial \phi}{\partial R} - j  k \phi \right) &= 0 \\
\\
\end{aligned}
\end{equation}
where $z=0$ is the waterline, $z=-h$ is the sea floor, $\Gamma$ is the body surface, $\vec{n}$ is the normal vector to that surface, and $\vec{u}$ is the body velocity and $k$ is the wavenumber. This BVP is linearized by assuming a small wave steepness and small motion of the WECs around a mean position. 
Similarly, the hydrodynamics module also calculates the hydrostatic stiffness matrix of the WECs ($C_{33}$). For a stably floating body $b$ a hydrodynamic analysis is then performed $(M_{bb} = \rho g V_b)$ .  The density of sea water is $\rho$, $g$ is acceleration due to gravity, $V_b$ is the submerged volume of the WEC, and $M_{bb}$ is the mass of the WEC.

Most common solvers modeling hydrodynamics are based on a reformulation of the governing PDE (Laplace equation) and associated boundary conditions as boundary integral equations (BIEs). These BIEs and the associated boundary element method (BEM) are well-suited for wave problems in unbounded domains and for complex geometries.  In this method, only the body geometry is discretized into a mesh with panels and Green's identity is used to derive a boundary integral equation which is then discretized into a linear system. Thus, this method is ideal for design (size and shape) optimizations as only the body surface needs to be meshed. This is advantageous compared to meshing the domain (large ocean) around the floating body. This method is implemented in the open-source frequency-domain BEM code Capytaine (\cite{ancellin_capytaine_2019} \cite{babarit_theoretical_2015}), our solver of choice. 

 When co-locating a system of interacting bodies, the energy produced when analyzed using linear potential flow theory may not linearly scale with the number of interacting bodies and has been shown to differ for various configurations from an isolated body (\cite{budal}). The hydrodynamic interactions due to the superposition of diffracted and radiated waves from oscillating bodies can constructively or destructively affect the motion. These interactions are a function of WEC geometry and array configuration and affect wave power absorption of each body in the arrays. These interactions are important to consider in early design studies.  Many different methods exists to simulate the array hydrodynamics such as analytical \citep{simon1982multiple}, semi-analytical (\cite{zhong2016wave}, \cite{mcnatt2015novel}, \cite{singh2013hydrodynamic}), predictive methods \citep{ZHU2022112072}, and full BEM resolution (interaction between all panels).
 Frequency domain BEM codes can solve the diffraction and radiation problems of wave-body interaction and calculate respective velocity potentials ($\phi_{\text{diffraction}}, \phi_{\text{radiation}}$), giving the total potential as

 \begin{equation}
     \phi_{\text{total}} = \phi_{\text{radiation}} + \phi_{\text{diffraction}} + \phi_{\text{incident}}
 \end{equation}

\begin{figure}[H]
    \centering
    \begin{subfigure}{0.45\textwidth}
        \centering
        \includegraphics[width=\textwidth]{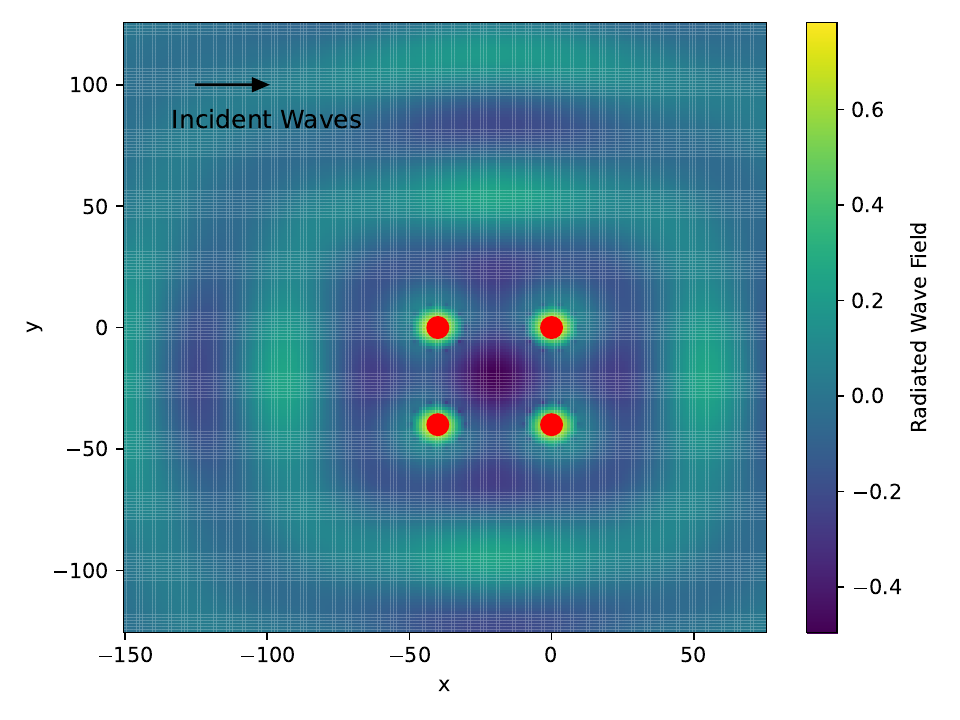}
        \caption{The radiated wave elevation.}
        \label{fig:rad}
    \end{subfigure}
    \hfill
    \begin{subfigure}{0.45\textwidth}
        \centering
        \includegraphics[width=\textwidth]{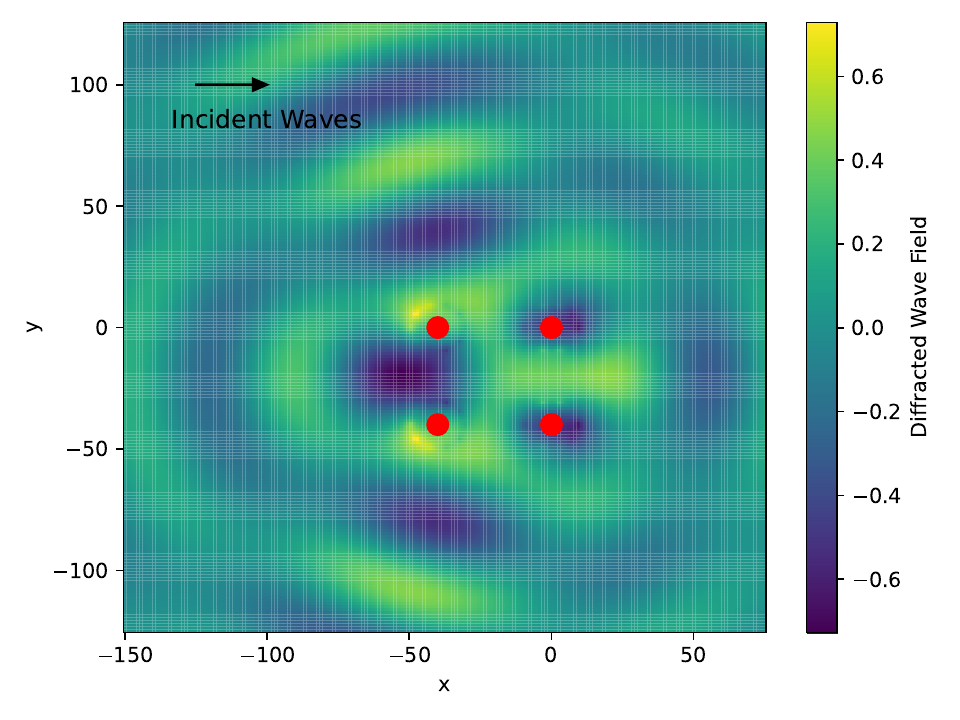}
        \caption{The diffracted wave elevation.}
        \label{fig:diff}
    \end{subfigure}
    \hfill
    \begin{subfigure}{0.45\textwidth}
        \centering
        \includegraphics[width=\textwidth]{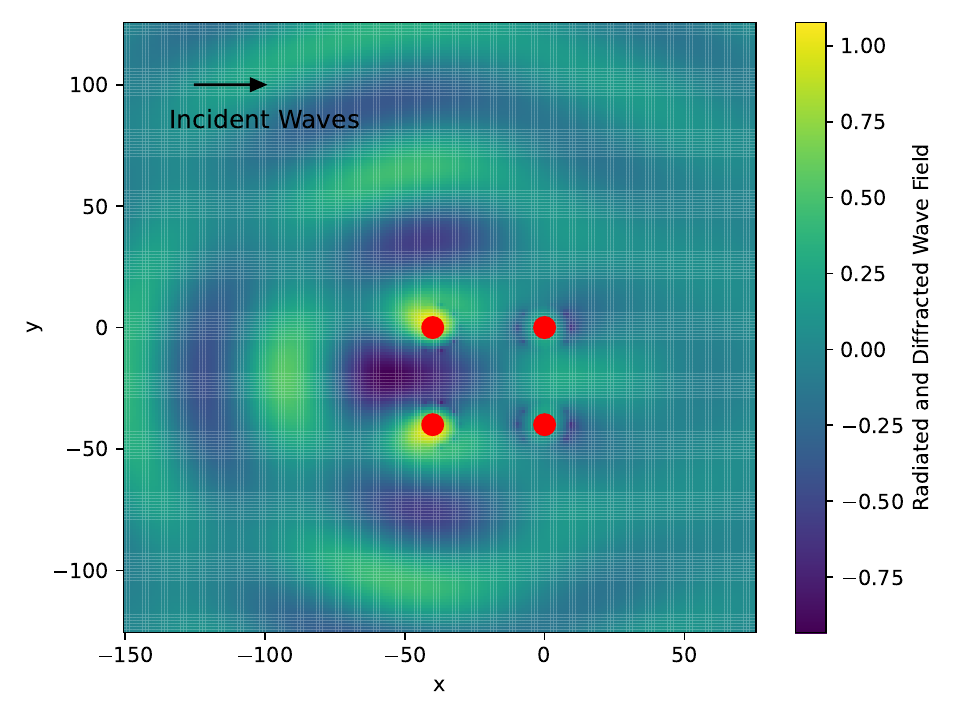}
        \caption{Both radiated and diffracted wave elevation.}
        \label{fig:rad_dif}
    \end{subfigure}
    \hfill
    \begin{subfigure}{0.45\textwidth}
        \centering
        \includegraphics[width=\textwidth]{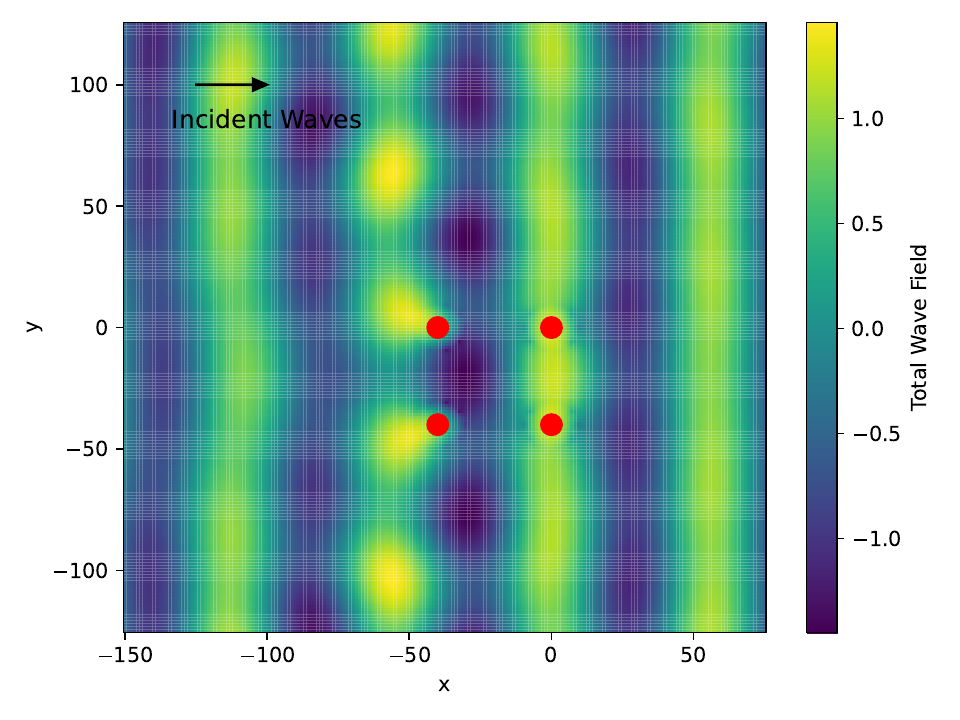}
        \caption{The total wave elevation.}
        \label{fig:total}
    \end{subfigure}
    \caption{The elevation of the radiated, diffracted, and total wave fields for an arbitrary 4-body WEC array.}
    \label{fig:wave elevation}
\end{figure}

The wave elevation of the radiated, diffracted, and total wave fields is shown in Fig. \ref{fig:wave elevation}, providing a visual for the wave-body interaction. The figure displays an arbitrary array of 4 WECs and how each component (radiation and diffraction) of the total potential for each WEC body impacts the surrounding wave field. Generally, the magnitude of radiated waves is slightly smaller than the diffracted waves, but both play a part in how the bodies oscillate.
 
$\phi_{\text{radiation}}$ is used to calculate the hydrodynamic force coefficients $\mathbf{A}_{4 \times 4}$ and $\mathbf{B}_{4 \times 4}$ (added mass and damping in heave). Added mass($\mathbf{A}$) and damping ($\mathbf{B}$) arise due to the motion of WECs. They are proportional to acceleration and velocity of the WEC respectively. They both depend on the wave frequency ($\omega$). Heave excitation force ($\vec{F}_{1\times 4}$) is found from the hydrodynamic pressure due to incident and scattered waves, the added mass, and the damping terms, described by

 \begin{equation} \label{eq:2}
     \vec{F} = -\omega \rho \int \int_{\Gamma} n_i \phi ds
 \end{equation}
 
  \begin{equation}
     \mathbf{A}_{ij} - \frac{j}{\omega}\mathbf{B}_{ij} = - \rho \int \int_{\Gamma} n_i \phi_{} ds
 \end{equation}

where $\omega$ is the wave frequency, $\rho$ is the fluid density, $n_i$ is the normal vector, and $\phi$ is the potential.

\subsubsection{Mesh Convergence}
 In order to ensure the accuracy of the coefficients, a mesh convergence study (see \ref{fig:mesh_convergence}) was done using few representative design obtained from Latin hypercube sampling ($N_S$ = 6) of the design space. The mesh resolution for the analysis is shown in the Table \ref{tab:resolution}.
 
\begin{table}[h]
\centering
\begin{tabular}{|c|c|}
\hline
 Radius (r) & Resolution (nr, n$\theta$, nx)         \\
\hline
 $> 5$        & $(7, 35, 25)$         \\
 $4 < r \leq 5$  & $(5, 25, 20)$         \\
 $3 < r \leq 4$  & $(3, 15, 15)$         \\
$\leq 3$        & $(2,10,10)$\\
\hline
\end{tabular}
\caption{Mesh resolution based on cylinder radius. nr, n$\theta$, and nx represent the number of panels along a radius at the end, the number of panels around a slice, and the number of slices respectively.}
\label{tab:resolution}
\end{table}

Despite the high accuracy of full BEM resolution, researchers have been reluctant to use it for design optimization and sensitivity studies, mostly due to the computational limits. The direct numerical simulation has the computational cost of $O(N_{p}^2)$ for Green's function evaluation and to setup the dense influence matrices and $O(N_{p}^3)$ for the solution of linear system, with $N_p$ being the number of panels on all the WECs in an array. Free surface Greens function is also challenging to compute \citep{newman1985algorithms}. Mesh convergence (see Fig. \ref{fig:mesh_convergence}) can be performed to reduce the number of panel required by only meshing to the required resolution (see Table \ref{tab:resolution}). The computational cost and convergence of the BEM solvers also depends on the choice of linear solver and the dense influence matrix representation. 

\begin{figure}[H]
    \centering
     \includegraphics[width = 1.1\linewidth]{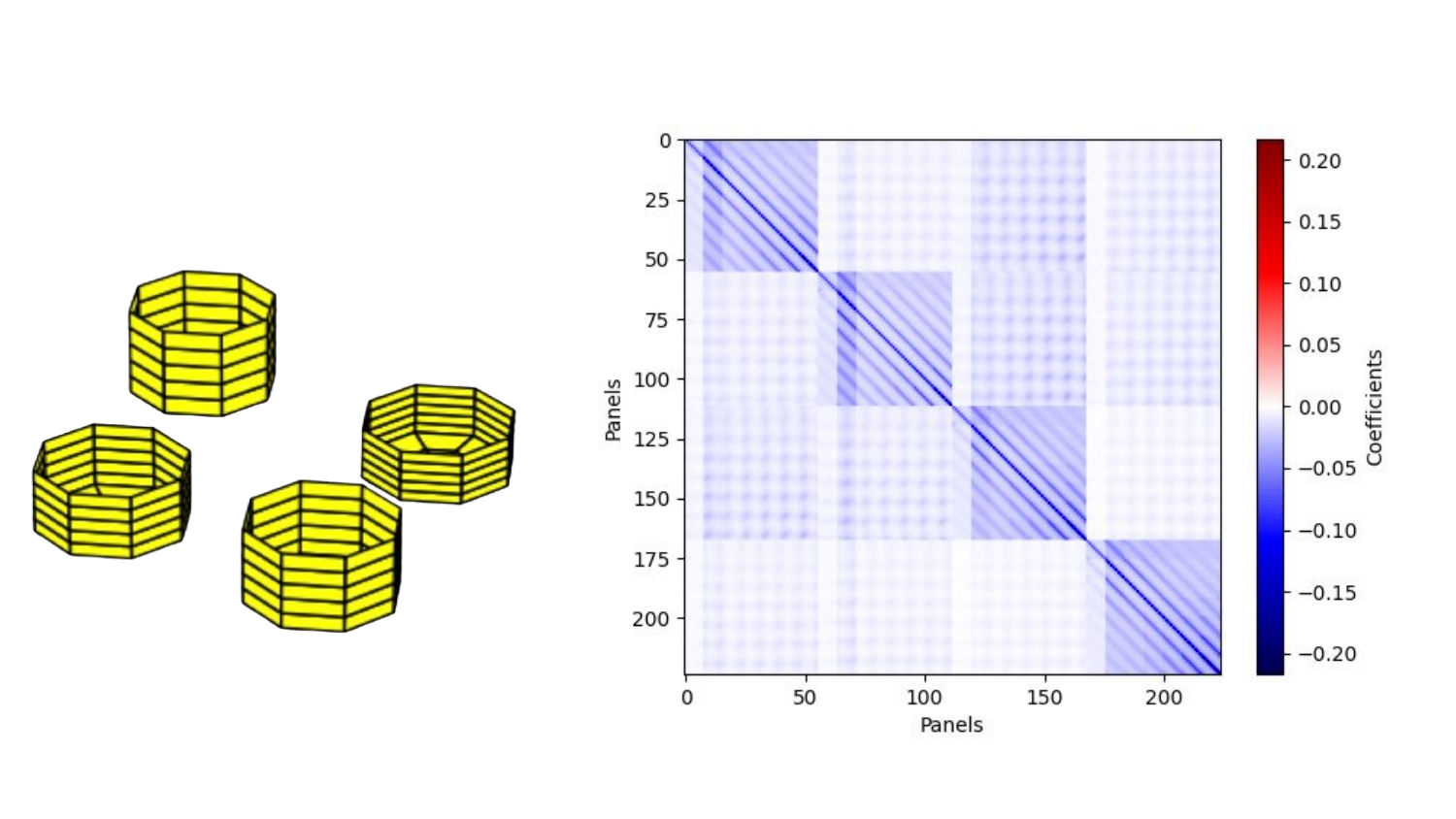} 
    \caption{Block structure in influence matrices computed via full BEM resolution for any design.}
    \label{fig:Hmatrices}
\end{figure}

With integration of modern methods in the BEM codes, the computational cost can be somewhat reduced allowing for accurate designs and sensitivity studies to be affordable. The limitation has been removed to some extent by integrating many modern methods from BEM literature into marine hydrodynamic BEM solvers \citep{Hmatrices_mathieu}. Each block in the influence matrices is the interaction between a group of panels (as illustrated on Fig. \ref{fig:Hmatrices}). The interaction between distant blocks can be approximated by a low-rank matrix. Adaptive cross approximation (ACA) can be used to construct this low-rank matrix. The tolerance and distance for ACA are fixed to be 1e-1 and 7$\times$radius, respectively.

Similarly, the boundary element solver has numerical issues such as irregular frequency and convergence issues. The removal techniques \citep{CHLee_removing} are not yet implemented in the open source numerical solver of choice - Capytaine. Thus, we add in a check to remedy this. A ``sniff test" is implemented to disregard the ill-conditioned influence matrices $(\kappa(A) \ge 500)$ to prevent the highly irregular added mass and damping coefficients. This is chosen based on our computational experiments although Newman suggests that BEM matrices build with free surface Greens function have condition number in  \textit{O(100)} assuming high accuracy in Greens function calculation \citep{newman1985algorithms}. If not accounted for, these can lead to erroneously high LCOE results. The genetic algorithm also avoids high LCOE results in its search for Pareto optimal designs. Any non-physical designs are further filtered during post processing.

\subsection{Dynamics, Controls, and Power Take-Off}
A simple controller commonly used in wave energy research is a reactive controller \citep{control}. Our WECs use a motor-actuated reactive controller, and any WEC, $i$, uses the design variable PTO damping coefficient, $d_i$, and a model determined PTO stiffness term, $k_i$, as the two controller gains. These two controller gains are used in motor-actuated controllers to determine the force on the motor ($F_{PTO}$) as follows
\begin{equation}
F_{PTO} = k\xi + d\frac{\partial \xi}{\partial t}
\label{control_force}
\end{equation}
where $\xi$ is the heave motion of the WEC.
For linear PTO devices, time-averaged power, $P$, for a given WEC, $i$, can be calculated with the following equation 
\begin{equation}
P_i = \frac{1}{2} d_i \big |j\omega \mathbb{E}_{i}(j\omega) \big |^2
\label{power}
\end{equation}
where $\mathbb{E}_i(j\omega)$ is the complex amplitude of heave motion. Heave motion is assumed to be sinusoidal for each WEC, and oscillating at same frequency of the waves, $\omega$. This allows the use of the following time domain equation
\begin{equation}
\xi_i = \mathbb{E}_i(j\omega)e^{-j\omega t}
\label{heave}
\end{equation}
where $t$ represents the time variable and $\xi_i$ is the time domain representation of heave motion. To calculate the complex amplitude of heave motion the coupled dynamics need to be solved. 
\begin{equation}
\frac{\vec{\mathbb{E}}(j\omega)}{A} = \big{[}-\omega^2(\mathbf{M}+\mathbf{A}) - j\omega(\mathbf{B}+\text{diag}(\vec{d})) + \mathbf{C} + \text{diag}(\vec{k})\big{]}^{-1} \vec{\mathbb{F}}(j\omega)
\label{dyn}
\end{equation}
The added mass matrix in heave ($\mathbf{A}$), hydrodynamic damping matrix in heave ($\mathbf{B}$), hydrostatic stiffness in heave ($\mathbf{C}$), and heave wave exciting force ($\vec{\mathbb{F}}(j\omega)$) all come from the hydro module. $A$ is the wave amplitude parameter, $\mathbf{M}$ is the diagonal array mass matrix, and $\text{diag}(\vec{d})$ forms a diagonal matrix using the values contained in the PTO damping vector. The components of the PTO stiffness vector ($\vec{k}$) are calculated to ensure decoupled resonance at the wave frequency $\omega$ using
\begin{equation}
k_i = \omega^2  (M_{ii} + A_{ii}) - C_{ii}
\label{stiff}
\end{equation}
where $M_{ii}$ is the mass of WEC $i$. 

Finally, the individual WECs power outputs are summed to obtain the total array power production. 

\subsubsection{Force Saturation}
With unconstrained reactive controllers tuned to resonate, it is common to see unrealistic forces on the PTO. To properly consider controller design, it is important to both consider controller gains as well as constraints. For example, any realization of a motor actuated controller will have some maximum force. Using the heave motion calculated by Equation \ref{dyn}, the total unsaturated force on each PTO, $\mathbb{F}_{\text{PTO},i}(j\omega)$ can then be calculated using 
\begin{equation}
\mathbb{F}_{\text{PTO},i}(j\omega) = \mathbb{E}_i(j\omega)\sqrt{k_i^2 + \omega^2 d_i^2}
\label{F_PTO}
\end{equation}
If the magnitude of this force is larger than the maximum PTO force parameter, $F_{max}$, the response needs to be saturated. The method used to saturate the response is the same as the method used by \cite{mccabe_multidisciplinary_2022}. Using that method, a multiplier, $\mu_i$, is found that modifies the PTO damping and stiffness for each WEC individually. If the magnitude of $\mathbb{F}_{\text{PTO},i}(j\omega)$ is less than or equal to $F_{max}$ then $\mu_i$ is set to 1 for that WEC.
\begin{equation}
\begin{aligned}
k_{i,sat} = \mu_i k_{i}\\
d_{i,sat} = \mu_i d_{i}
\label{new_PTO}
\end{aligned}
\end{equation}
These new saturated PTO gains can be used to recalculate motion ($\vec{\mathbb{E}}(j\omega)$) using Equation \ref{dyn}, with $\vec{k}_{sat}$ and $\vec{d}_{sat}$ substituted for $\vec{k}$ and $\vec{d}$. This recalculated heave is then used in Equation \ref{power} to calculate power.

It is worth noting that the force saturation model does not fully account for the coupling due to the array interactions. The motion of one WEC impacts the motion of all other WECs, meaning that performing the force saturation procedure on any one WEC the PTO force ($\mathbb{F}_{\text{PTO},i}(j\omega)$) changes for all the others. Another method for handling the force saturation that also fails to fully address coupling is to perform the saturation method one WEC at a time, recalculating the heave motion for all WECs after each saturation. A comparison of the results of these two methods (with $F_{max}$ set to $10^5$ N) is shown in Table \ref{tbl:sat}. The table shows that there is minimal, ~5\%, difference between the two methods, and was used to justify the the force saturation method we used. Future work is required to fully capture the coupling effects in the force saturation model within the larger context of the multidisciplinary model.

\begin{table}[]
\caption{Comparison of the two different force saturation modeling methods discussed. Method 1 is the method used in the study that saturates all the WECs at once. Method 2 saturates the WECs one at a time. The Balitsky column numbers come from an array like the one pictured in Fig. \ref{fig:balit_layout} in the Appendix \citep{valid}, and the diamond layouts are pictured in Fig. \ref{fig:diamond} in the Appendix. The WECs used in the Balitsky layout have a radius of 5 m, length of 4 m, and damping coefficient of 3.6$\times 10^5$ Ns/m. The WECs used in the diamond layouts have a radius of 7 m, length of 0.7 m, and a damping coefficient of 5.01$\times 10^5$ Ns/m.}
\begin{tabular}{llll}
 & Balitsky & Diamond (tight) & Diamond (spacious) \\ \hline
No Saturation & 1328.2 kW & 516.0 kW & 970.5 kW \\
Method 1 & 507.9 kW & 191.9 kW & 150.3 kW \\
Method 2 & 505.9 kW & 166.1 kW & 135.4 kW \\
Saturation change & -62\% & -63\% & -85\% \\
Method 2 change & -0\% & -5\% & -2\%
\end{tabular}
\label{tbl:sat}
\end{table}

\subsubsection{Q-factor}
We extend the discussion on the optimal average power with the q-factor. Q-factor ($q$) is commonly used to evaluate array power output. This metric is the ratio of the power produced by an array configuration of $n$ bodies ($P_{array}$) to the power produced by $n$ isolated bodies, shown by
\begin{equation}
    q = \frac{P_{array}}{n \times P_{isolated}}
\end{equation}


\subsection{Economics}\label{sec:econ}
The economics module serves to compute the LCOE

\begin{equation}
    \text{LCOE} = \frac{\text{CAPEX}_{ann} + \text{OPEX}}{\text{AEP}}
\end{equation}

where $\text{CAPEX}_{ann}$ is the annualized capital expenses, OPEX is the annual operating expenses, and AEP is the annual energy production. $\text{CAPEX}_{ann}$ includes the cost of each WEC based on mass, the cost of the mooring lines, the installation cost, and the shipping costs. OPEX incorporates maintenance costs, mid-life refit costs, and decommissioning costs (all annualized). The median $\text{CAPEX}_{ann}$ values and method for finding OPEX reported by the OES (\cite{chozas2015international}, \cite{chang2018comprehensive}) are used in this calculation and are scaled by the mass of the WEC

\begin{equation}
    \text{xPEX} = \text{xPEX}_{med} + \Bigg ( \text{xPEX}_{med} * \frac{m}{m_{max}} \Bigg )
\end{equation}

where xPEX represents either CAPEX (pre-annualized) or OPEX and $m_{max}$ is the mass of the largest WEC included in the OES report. The reported xPEX values are normalized by the rated power, so they must be multiplied by the rated power of the corresponding WEC and then scaled by the modeled WECs. The median CAPEX value is taken directly from the EOS report as \$9000/kW, and the median OPEX is taken to be 5\% of the median CAPEX, or \$450/kW. The CAPEX is annualized using

\begin{equation}
    \text{CAPEX}_{ann} = \text{CAPEX} \frac{i(1+i)^L}{(1+i)^L - 1}
\end{equation}

where $i$ is the interest rate and $L$ is the WEC lifetime. The annual energy production is calculated by

\begin{equation}
    \text{AEP} = P \times \eta_{trans} \times \eta_{avail}
\end{equation}

where $\eta_{trans}$ and $\eta_{avail}$ account for losses due to transmission lines and availability of wave power, respectively. This method effectively finds the ratio between WEC mass and mechanical power production and reports the corresponding economics metric.

\section{Validation}
The hydrodynamics and dynamics modules along with the power production were validated against another study involving arrays of cylindrical WECs (\cite{valid}). The study analyzed the interactions between two clusters of WECs using coupled BEM and MILDwave model (MILDwave is a numerical wave propagation model). The layout of each cluster is shown in the Appendix in Fig. \ref{fig:balit_layout}. Each cylinder has a radius of 5 meters and a length of 4 meters, and uses resistive (meaning the PTO stiffness vector $k$ is all zeros) control with each WEC having a PTO damping gain of 0.360 MNs/m. Note that this study also does not use force saturation so that is not included in this validation. The incoming waves ($\beta = 0$) are going from left to right with a period of 6 seconds and a unit amplitude. The two clusters are placed directly downstream from each other with three different spacings between clusters (500 m, 1000 m, 2000 m). A comparison is shown in Table \ref{tbl:valid}. We see a maximum error of 1.55\% between this model's results and the results of Balitsky et al. 

\begin{table}
\centering
\caption{Comparison of model in this paper with results from Balitsky et al.}
\begin{tabular}{llll}
Cluster Spacing & Balitsky & This Paper & Error \\ \hline
500 m & 1381 kW & 1392 kW & +0.80\% \\
1000 m & 1682 kW & 1707 kW & +1.55\% \\
2000 m & 1595 kW & 1589 kW & -0.38\%
\end{tabular}
\label{tbl:valid}
\end{table}

\section{Results}
\subsection{Pareto}
\begin{figure}[H]
    \centering
    \includegraphics[width = 0.7\linewidth]{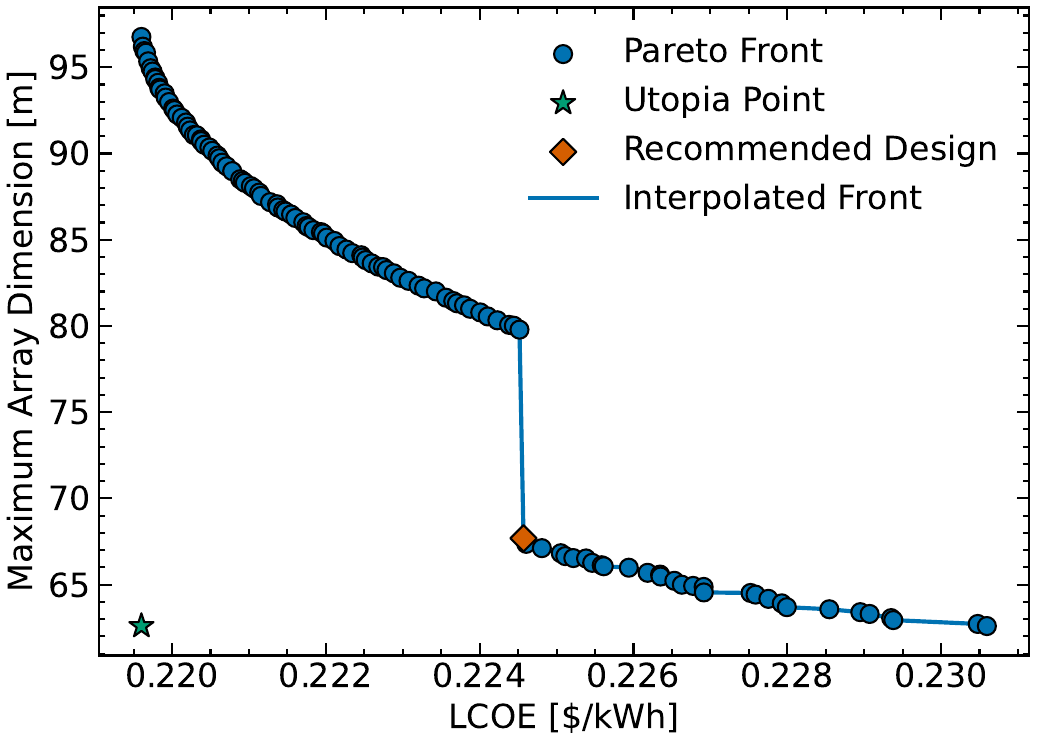}
    \caption{Pareto front generated by NSGA-II. A `recommended design' (shown in Fig. \ref{fig:designrec}) is highlighted for which further sensitivity analysis is performed. The decision space looks discretized across the pareto front.}
    \label{fig:pareto}
\end{figure}

\begin{figure}[H]
    \centering
    \begin{subfigure}{0.45\textwidth}
        \centering
        \includegraphics[width=\textwidth]{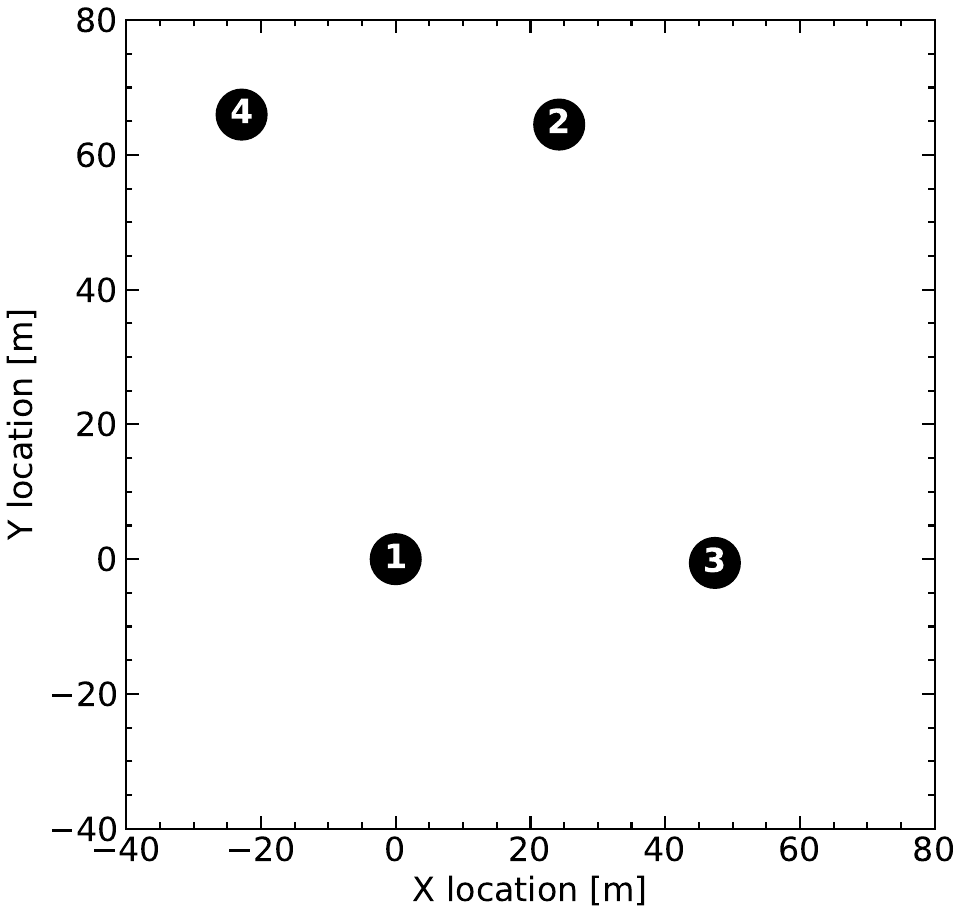}
        \caption{Minimum LCOE design}
        \label{fig:designlcoe}
    \end{subfigure}
    \hfill
    \begin{subfigure}{0.45\textwidth}
        \centering
        \includegraphics[width=\textwidth]{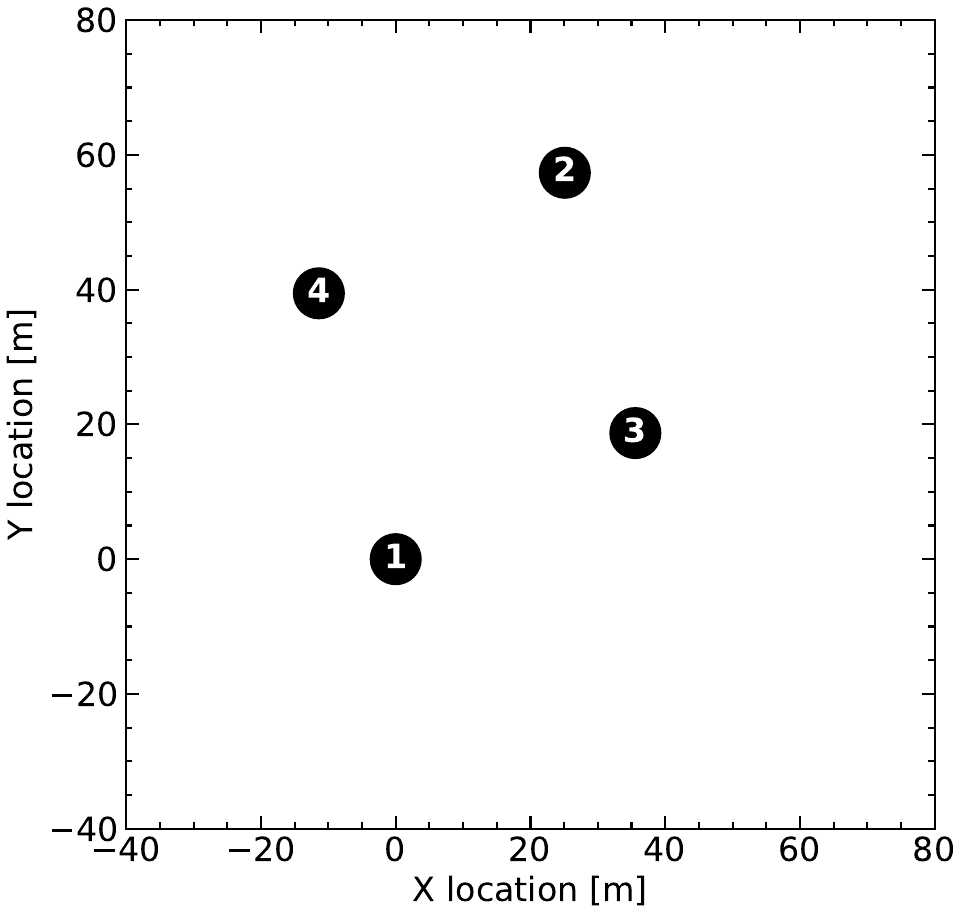}
        \caption{Minimum ocean space design}
        \label{fig:designspace}
    \end{subfigure}
    \caption{Minimum LCOE and ocean space designs from the Pareto front. }
    \label{fig:designs}
\end{figure}

The Pareto front, resulting from the optimization in Section \ref{opt} shown in Fig. \ref{fig:pareto}, depicts the trade-off between minimizing farm length and minimizing LCOE. To reach the lowest LCOE, a larger amount of ocean space is required; however, a reasonable LCOE can be still be achieved with a much smaller farm length with optimal radii, lengths, and damping coefficients. The lowest ocean space solution shows a 37\% reduction in ocean space used compared to the minimized LCOE; with only a 5\% increase in LCOE. The mean damping for the minimal LCOE design is 357 kNs/m, with a 0.0130 coefficient of variation. The mean damping for the minimal spacing design is 369 kNs/m, with a 0.0480 coefficient of variation. The minimized LCOE configuration can be seen in Fig. \ref{fig:designlcoe}, while the minimized farm length configuration can be seen in Fig. \ref{fig:designspace}. The distance between them vary depending on their size; designs that are compact tend to have higher LCOE values than designs spaced far enough apart such that interactions are negligible. The resulting rhombus-like configuration, at varying levels of compactness, is optimal for four WEC arrays in terms of minimizing both LCOE and ocean space. However, more compact configurations may require more individualized controls in order to reduce energy losses. Note that some of the designs found through the optimizer were removed in post processing due to the nonphysical behavior predicted by our model. The pareto front with this points included can be found in the Appendix Fig. \ref{fig:pareto_full}.

\begin{figure}[H]
    \centering
    \includegraphics[width = 0.7\linewidth]{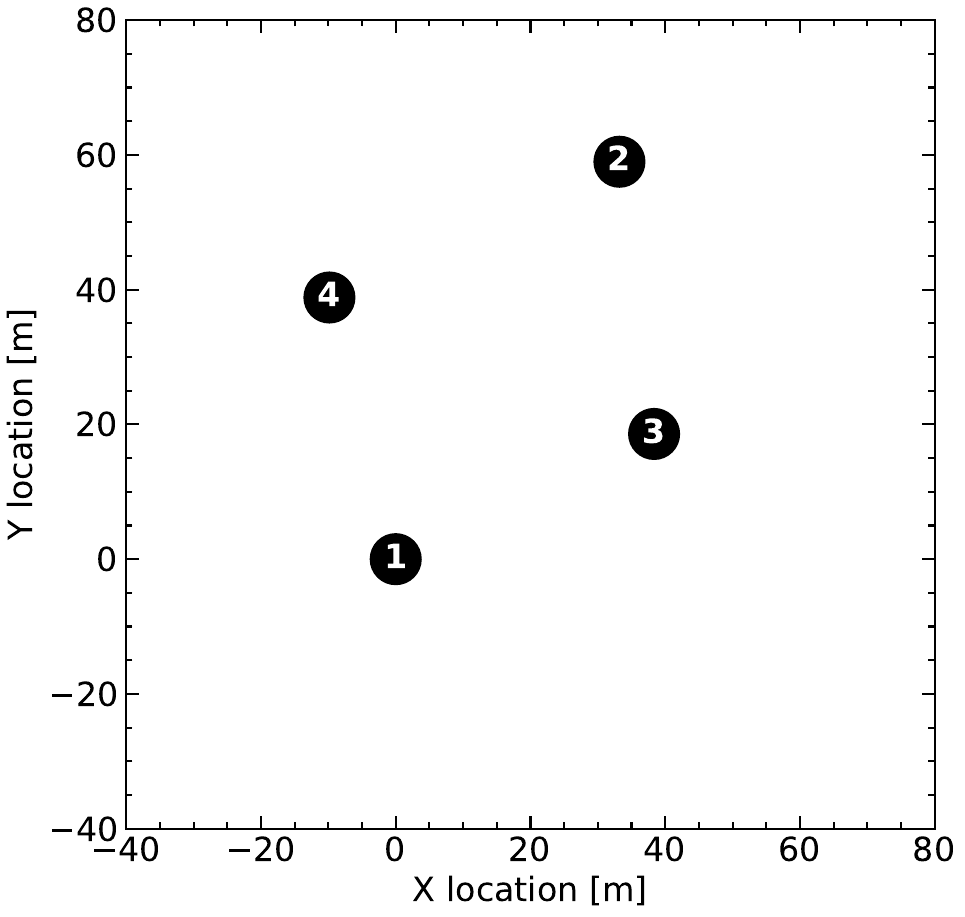}
    \caption{Recommended design}
    \label{fig:designrec}
\end{figure}

\begin{figure}[H]
    \centering
        \centering
        \includegraphics[width=0.6\textwidth]{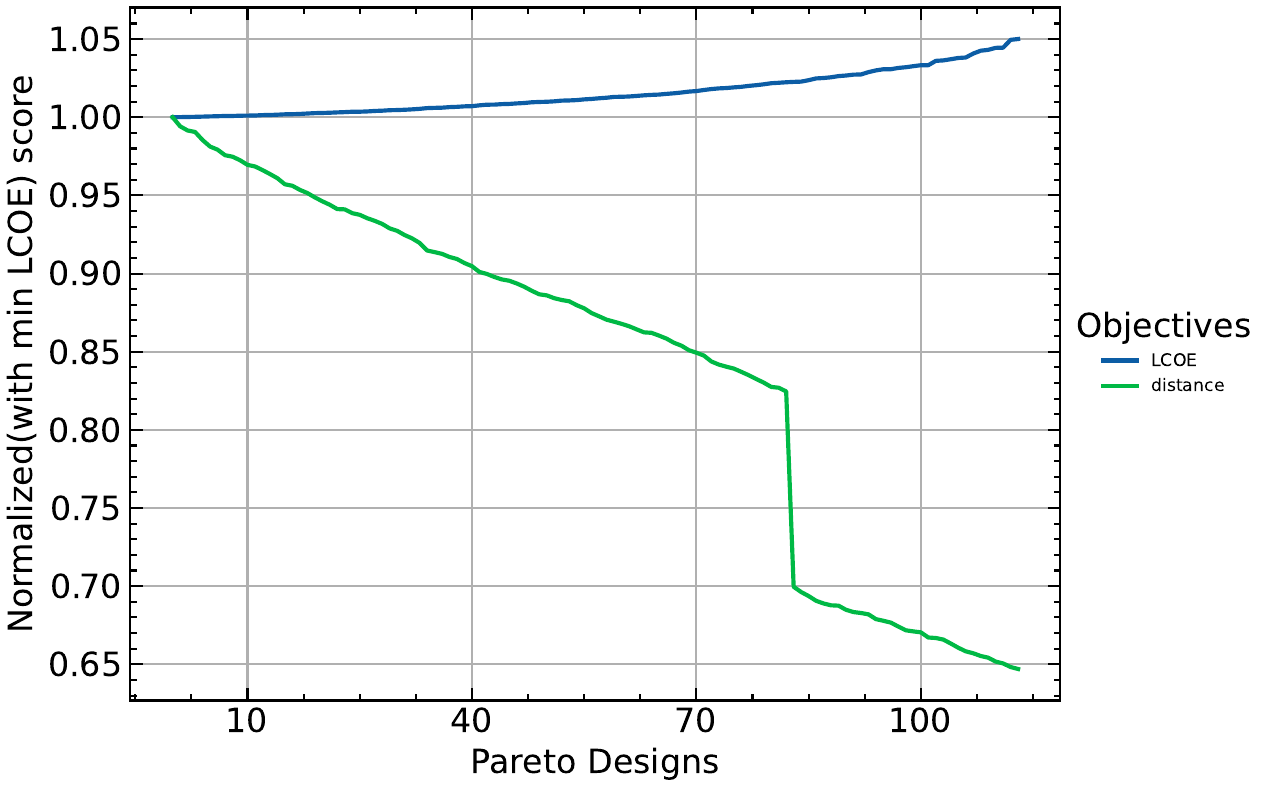}
        \caption{Change in the objectives across the pareto design normalized with min LCOE design }
        \label{fig:percent_designlcoe}

\end{figure}

\subsection {Disturbance Coefficients}
\begin{figure}[H]
    \centering
    \begin{subfigure}{0.45\textwidth}
        \centering
        \includegraphics[width=\textwidth]{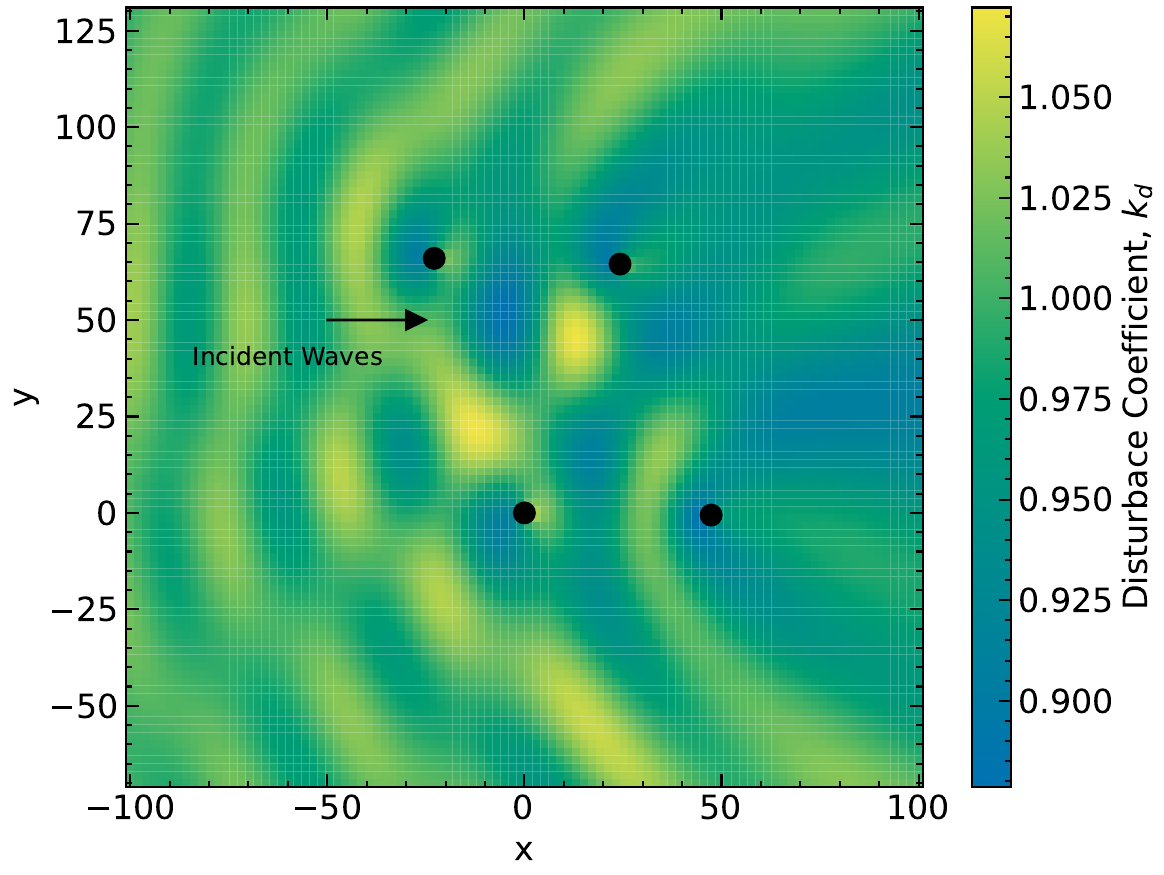}
        \caption{Minimum LCOE design}
        \label{fig:kdlcoe}
    \end{subfigure}
    \hfill
    \begin{subfigure}{0.45\textwidth}
        \centering
        \includegraphics[width=\textwidth]{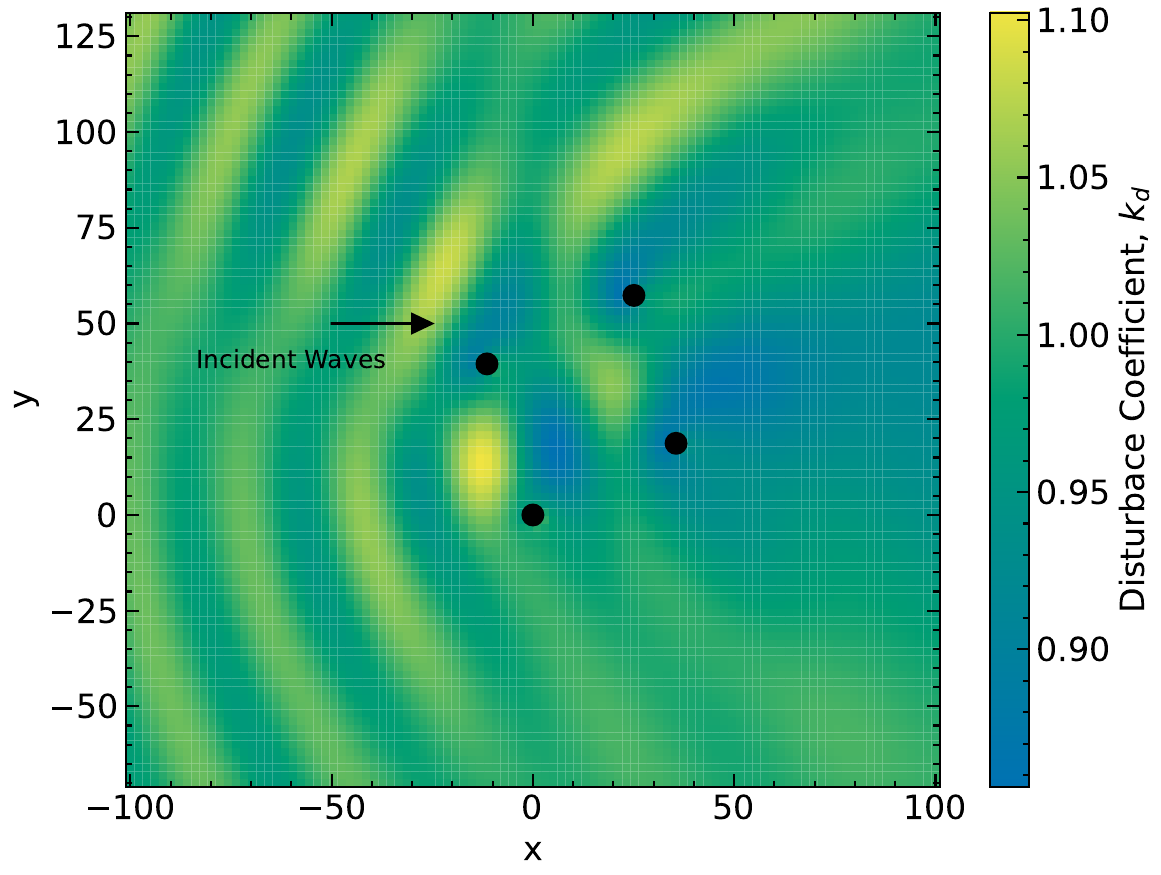}
        \caption{Minimum ocean space design}
        \label{fig:kdspace}
    \end{subfigure}
    \caption{Disturbance coefficient ($k_d$) for the wave fields around two pareto designs.}
    \label{fig:kd}
\end{figure}

 The disturbance coefficient is the ratio of the total wave elevation ($\zeta_{total}$) in the presence of WECs to the incident wave elevation ($\zeta_{inc}$), defined as
 \begin{equation}
     k_d = \frac{\zeta_{total}}{\zeta_{inc}}
 \end{equation}
This ratio shows the magnitude of which the presence of the bodies affects the surrounding wave field. Fig. \ref{fig:kd} shows the disturbance coefficient ($k_d$) for the minimized LCOE and minimized farm length configurations. The higher $k_d$ values in front of and to the side of the WECs indicate higher wave reflection from those bodies. The amount of wave height a WEC body can reflect directly relates to how much power they are able to extract from the waves. Additionally, the configuration of the WECs leads to higher reflections for each body due to the interaction effects. The magnitude of these reflections is comparable for both configurations; however, upstream of the minimized LCOE configuration exhibits more increased wave height overall. This indicates more synergistic effects than the minimized farm length configuration. In both configurations, WECs furthest from the incoming wave front experience shadowing effects, where the downstream wave height is not as great as upstream. However, these effects are minimal in our Pareto optimal solutions compared to a standard grid or un-optimized staggered layout.
 
\subsection{Q-Factors}
\begin{figure}[H]
    \centering
    \begin{subfigure}{0.45\textwidth}
        \centering
        \includegraphics[width=\textwidth]{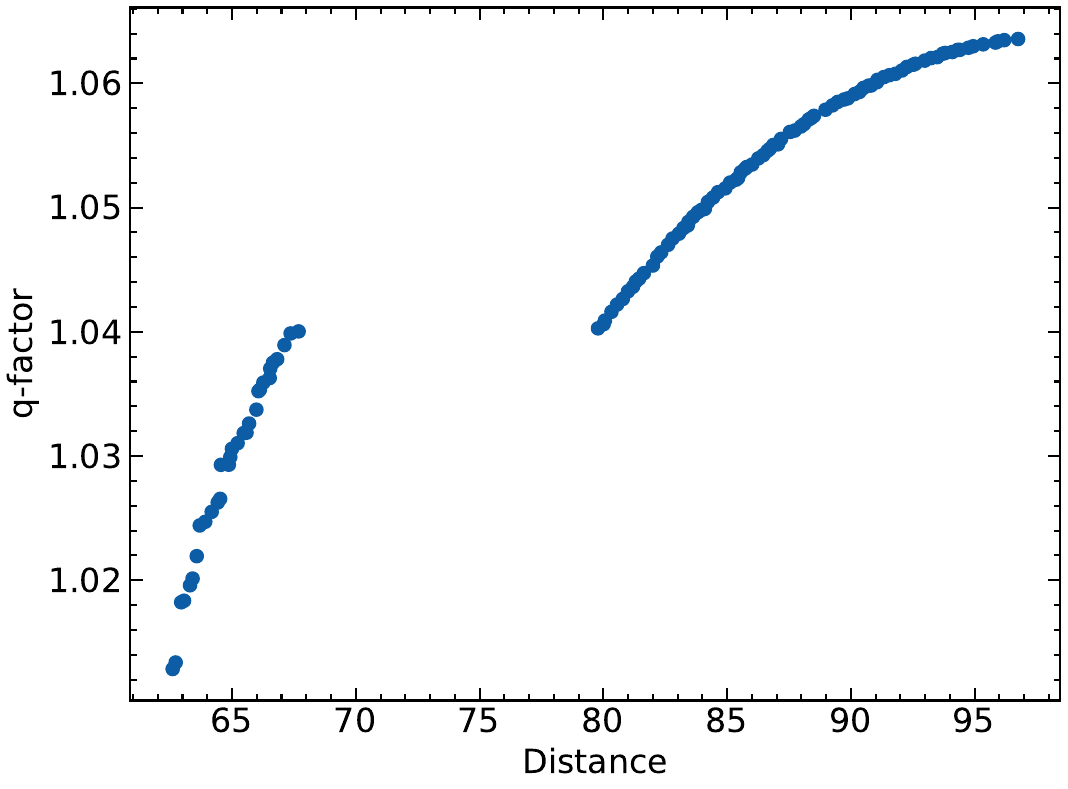}
        \caption{q-factor vs. dist}
        \label{fig:min_ocean}
    \end{subfigure}
    \hfill
    \begin{subfigure}{0.45\textwidth}
        \centering
        \includegraphics[width=\textwidth]{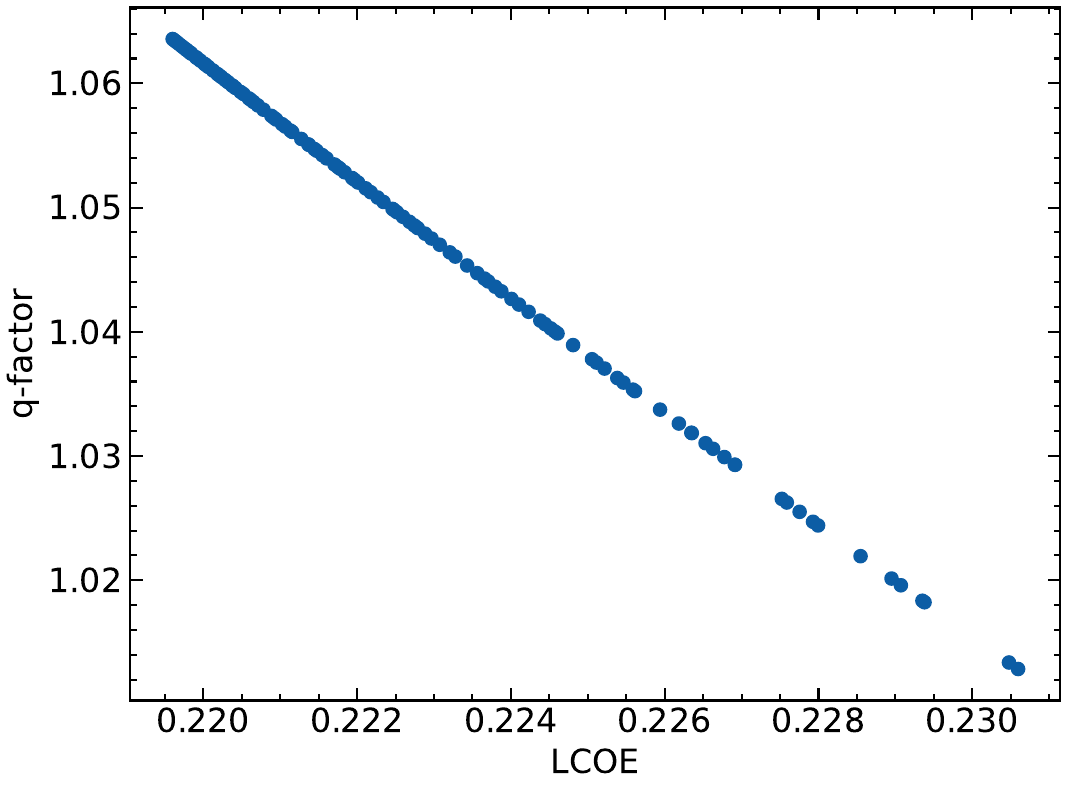}
        \caption{q-factor vs LCOE}
        \label{fig:min_lcoe}
    \end{subfigure}
    \caption{Variation of q-factor for each objective for the set of optimal designs}
    \label{fig:qfactors}
\end{figure}

The q-factors for the Pareto front designs are shown in Fig.  \ref{fig:min_ocean} and \ref{fig:min_lcoe}. Favorable q-factors are found for all designs, showing that these Pareto designs out-preform isolated WECs. The cluster on the Pareto front with the minimal LCOEs reach the highest q-factors, up to 1.06. This confirms the way the bodies interact in that specific configuration slightly magnifies the power production. The extent which the amplification is felt could potentially be under-predicted by mid fidelity model such as BEM, but still leads to solid array designs. The interaction of the fidelity of the model and controls parameters is a topic for future work.

\subsection{Ad-hoc Regression Model For Trade-off Analysis}
\begin{figure}[H]
    \centering
\includegraphics[width=\linewidth]{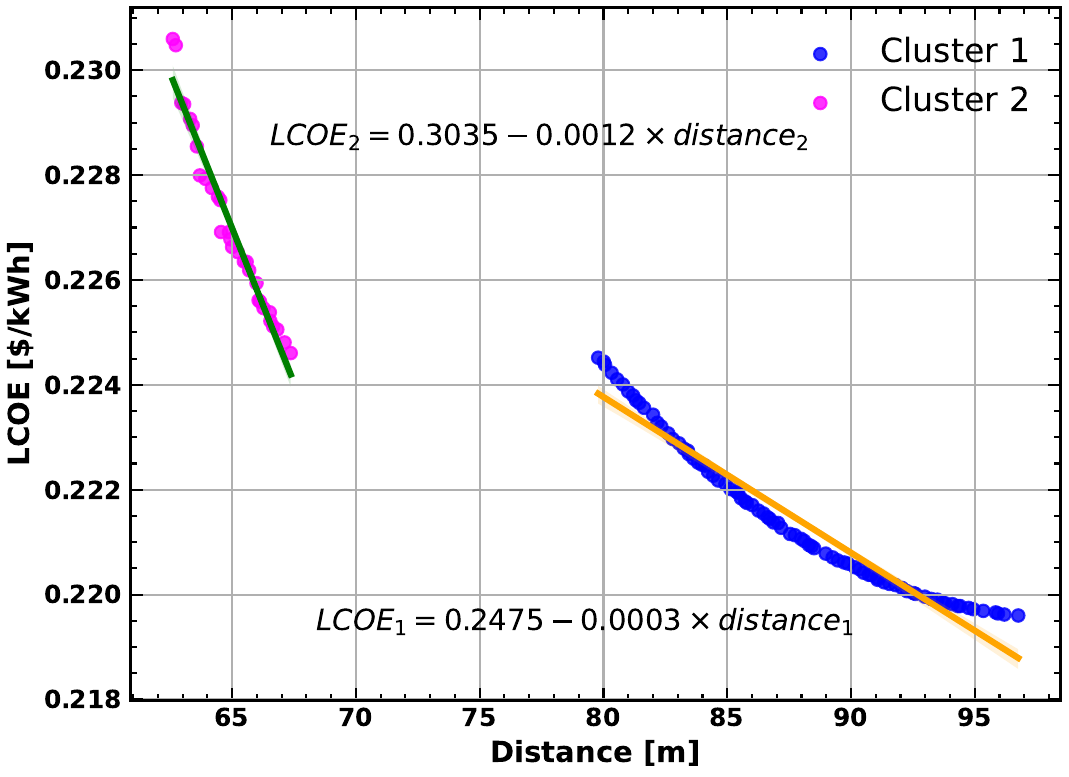}
    \caption{Different linear models for tradeoff analysis in discrete design space in the Pareto front from the multiobjective optimization.}
    \label{fig:regressionmodel}
\end{figure}

The interaction of the objectives and design variables in the Pareto front are analysed for system level inferences. A simple regression model (see Fig. \ref{fig:regressionmodel} is fitted for interpret-ability for the relation between the objectives (LCOE and distance) for the discrete decision space as shown by the Pareto front clusters (see Fig. \ref{fig:pareto}). Analysis like this are essential to help facilitate the stakeholders and managers understand and interpret the results of multi-objective problems. Both models explain around 95\% ($[0.95,0.96]$) variation expected in the LCOE for optimal designs. The expected LCOE is shown by
\begin{align}
    LCOE_{1} &= 0.2475 - 0.0003 \times distance_{1} \\
    LCOE_{2} &= 0.3035 - 0.0012 \times distance_{2}
\end{align}

The subscripts indicate clusters. $1$ is the cluster with low LCOE and $2$ is the cluster with low maximum array dimensions. The optimal (minimized) LCOE in compact layouts can be expected to decrease by 3e-4 (\$/kWh) for every meter increase in the distance and by 1.2e-3 (\$/kWh) for expanded arrays. The relationship in the larger layouts is not as linear as the decision in compact arrays. It is discussed here strictly for interpret-ability purposes rather than predictive purposes. This analysis is however not extensible to other optimization where the constraints, bounds and number of design variable are different.

\section{Global Sensitivity Analysis}

Sobol Sequence method is used to generate the samples using Python package SA-Lib (\cite{salib}) over a multi-dimension parameter space from different technical domain outlined in xDSM (see Fig. \ref{fig:xdsm}) diagram. The goal is to fully capture the relationship between the unconditional variance  of optimal WEC farm design to the variance of the dominant parameters in each of the involved disciplines. 
A convergence study is first performed to assess the required sampling size as the accuracy of this method depends the sample size. There is a limiting benefit to the sample size.  Convergence around $N_{sobol} = 4000$ for total samples $ = N_{sobol} \times (2D + 2)$ suggests that is good enough samples for sensitivity calculations for most parameters (see Fig. \ref{fig:sobolConv}). $D$ being the total number of parameter inputs $(D= 7)$ and using 1000 bootstrapping samples to get the average of the sensitivities. 
 
 Sobol variance decomposition method expresses the variation of the output as the superposition of the variation of the input. The total variance $V$ of the model output $Y$ can be decomposed as

\begin{equation}
V(Y) = \sum_{i=1}^{d} V_i + \sum_{i<j}^{d} V_{ij} + \ldots + \sum_{i<j<k}^{d}V_{1\ldots d}
\end{equation}

where $V_i$ is first-order or independent effect of the parameter,  $V_{ij}$ is second-order effect, and $d$ is the total number of input parameters. 

Sobol indices provide insights into the how the uncertainties of the input parameters and their interactions contributes to the variance of the output (minimized LCOE). For each parameter ($X_i$), the total Sobol index gives a measure of total influence on ($LCOE$) along with its interaction with other parameters ($X_j$). This allows for an assessment of the subsystems parameter interactions on the Pareto optimal designs. The total sensitivity of a design ($ST$) is

\begin{equation}
    ST = 1 - \frac{V(E(Y|X_i))}{V(Y)}
\end{equation}
The total sensitivity for a design in the Pareto set with design vector for recommended design (see Figs. \ref{fig:pareto} and \ref{fig:designrec}).

\begin{table}[htbp]
\centering
\caption{Parameter Bounds. All parameters are uniformly sampled in the bounds}
\label{tab:param_problem}
\resizebox{\textwidth}{!}{%
\begin{tabular}{|c|c|c|c|c|}
\hline
Parameter & Description & Lower Bound & Upper Bound & Total Sensitivity\\
\hline
wave\_heading & Wave Direction (rad) & -$\pi$ & $\pi$ & 0.1 \\
omega & Wave Frequency (rad/s) &0.1 & 3 & 0.04\\
$L$ & WEC Lifetime (years)& 5 & 35 &3.0e-3\\
wave\_amplitude & Amplitude of waves (m)& 0.2 & 3 &1e-2\\
interest & Interest rate (-)&0.05 & 0.2 &3e-4 \\
$n\_avail$ & Wave availability (-)&0.79 & 0.99 &1.5e-4\\
$S$ & Array Scaling Factor (-)& 0.5 & 0.99 &7.7e-5 \\
$F_{max}$ & Maximum saturation force (N)&1e4 & 1e6 &1.5e-8\\
\hline
\end{tabular}}
\end{table}

 Wave heading variance impacts about 10\% of the total variance of LCOE suggesting accurately modeling wave heading in the hydrodynamic analysis of WEC farms increases the reliability. Wave heading being sensitive is expected because although a single point absorber WEC is axisymmetric, the array is not. Thus, designers aiming to reduce LCOE by 10\% should try to reduce all variance of wave heading. Similarly, all others parameters with very small sensitivity index can be set at a constant value for the multidisciplinary system analysis. This confirms with what is generally expected for `linear' models in each of the subsystem. From the multidisciplinary optimization point of view, we confirm the change in the wave environment, interest rate, and lifetime have some combined impact on the variability of the LCOE of the WEC farms. The viability of the WEC farm obtained here will differ by wave environment more than economics parameter probably due to simplified economics module adopted in this optimization. Note that the hydrodynamics solver here assumes a regular wave and the wave environment parameters are all assumed independent. This could be different for the analysis under irregular waves and correlated parameters. As illustrated in the Table \ref{tab:param_problem}, the low sensitivity indices indicate that the problem is more or less linear. This suggests for system level analysis and optimization, the inclusion of more constraints, higher fidelity of governing equations solvers in each of the coupled discipline is needed to fully account the variance in the LCOE and reduce the risk.
\begin{figure}
    \centering
\includegraphics[width = 0.5\linewidth]{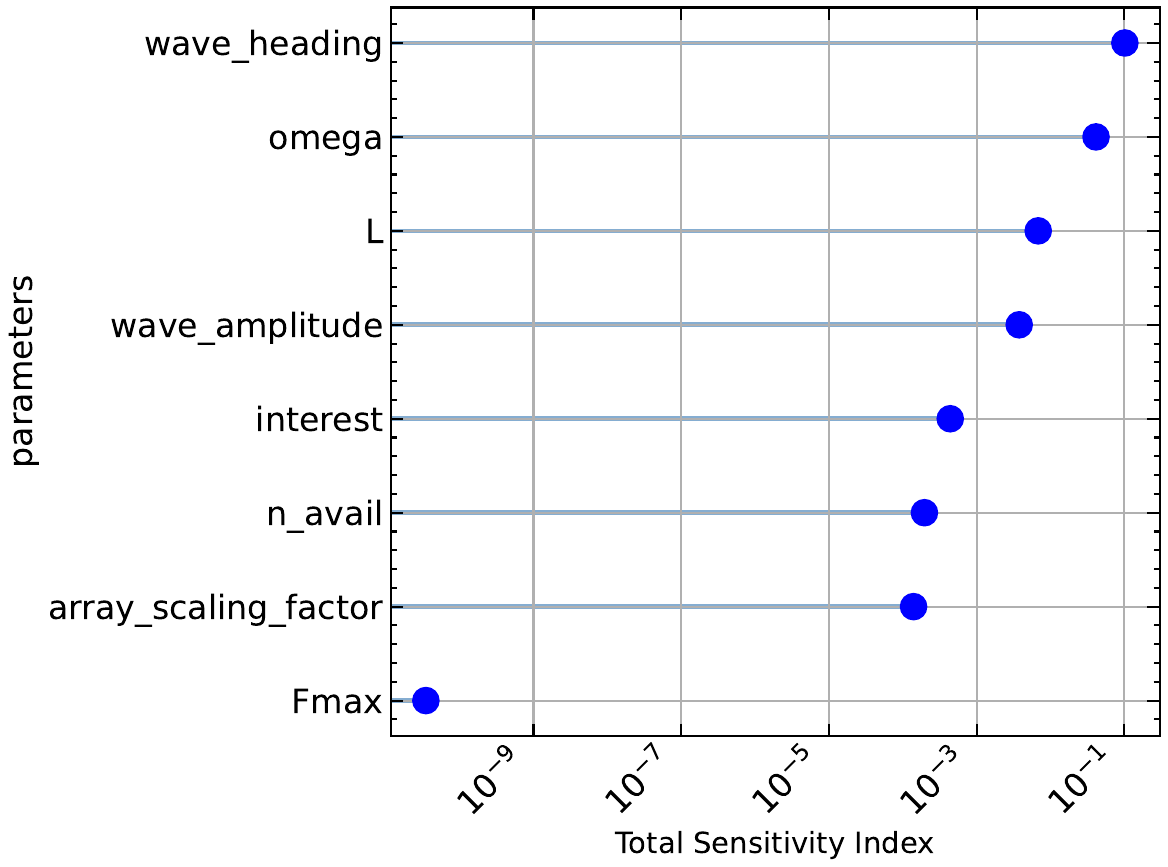}
    \caption{Total sensitivity (ST) of the optimal LCOE (design with LCOE = 0.21, distance =53.5) to parameters. Omega, wave heading and draft seems to be sensitive to the LCOE.}
    \label{fig:enter-label}
\end{figure}

\section{Future Work}
For the future work, similar analysis for large number of WECs with variable fidelity of the governing equations ($R_i(\mathbf{x},\mathbf{p};\hat{u}$)) should be included. It should be noted that the BEM computational cost increases with an increasing number of WECs, so there is a strong need to adopt modern BEM methods and efficient computational tools such as GPUs in the hydrodynamic analysis. The numerical issues in the existing open source solver such as convergence and irregular frequency also causes difficulties in the analysis and optimization. Mooring and transmission lines are not considered in the dynamics; however, future work should investigate the cost savings of sharing these lines, potentially making tighter configurations more favorable. Since this study only considers a single monochromatic wave, future studies should look at a JONSWAP spectrum or another similar spectrum that more accurately approximates a real sea-state. Not only should a frequency spectrum be considered in future work but, a spectrum of wave headings should also be considered. Additionally, an improved hydro module, or perhaps adding a feasibility constraint to match the solver fidelity, is necessary to prevent the optimizer from finding designs that lead to nonphysical results.

\section{Conclusion}
This paper details a multidisciplinary design optimization problem for WEC farms. A multidisciplinary analysis consisting of the array hydrodynamics, controls, and economics are discussed within this framework. For simplicity, only feed forward couplings were discussed. However, this framework can be extended to include the feedback coupling (for example, between mooring and hydrodynamics). A multi-objective genetic algorithm was used to analyze the relationship between competing objectives of minimizing farm length and minimizing LCOE. A sensitivity analysis of the Pareto optimal designs is conducted to evaluate their importance. The optimal rhombus-like configuration has varying levels of compactness as suggested by clusters in the Pareto fronts. This suggests there exists some configurations which are more `power dense' than the others. The designs with these trade-off should be studied more. It is shown that the optimal LCOE (minimized) is sensitive to the variance in wave environments, mostly wave heading ($\beta$) and wave frequency ($\omega$) thus providing information to designers on where to target to reduce risk (uncertainty). Similarly, a regression model is developed to understand the trade-off between the two objectives for different decision space in the Pareto front.

This paper adopts a multidisciplinary systems analysis and optimization methodology to provide guidance to both WEC array designers and to the stakeholders regarding Pareto-optimal design heuristics, sensitivities, model calibration and uncertainty (risk) assessments. 
The analysis and the optimization code in this paper are open-source and can be accessed in the github: \url{https://github.com/symbiotic-engineering/wec_array_opt}.

\section{CRediT authorship contribution statement}
\textbf{Kapil Khanal:} Conceptualization, Methodology, Software, Formal analysis, Investigation, Data Curation, Original Draft, Review \& Editing, Visualization. \textbf{Nate DeGoede:} Conceptualization, 
 Methodology, Software, Validation, Writing - Review \& Editing. \textbf{Olivia Vitale:} Conceptualization, Methodology, Writing - Review \& Editing, Visualization. \textbf{Dr.Maha N Haji:} Resources, Supervision, Writing - Review \& Editing, Funding acquisition

\section{Acknowledgement} 
Authors would like to thank 
Carlos S. Michelen, Matthieu Ancellin, Rebecca McCabe and Matthew Haefner for their valuable feedback on the simulation and analysis in this manuscript. This work was supported in part by the Graduate Fellowship (for first year doctoral students) via Systems Engineering Department.

\appendix

\section{Convergence}
\begin{figure}[H]
    \centering
\includegraphics[width = 0.7\linewidth]{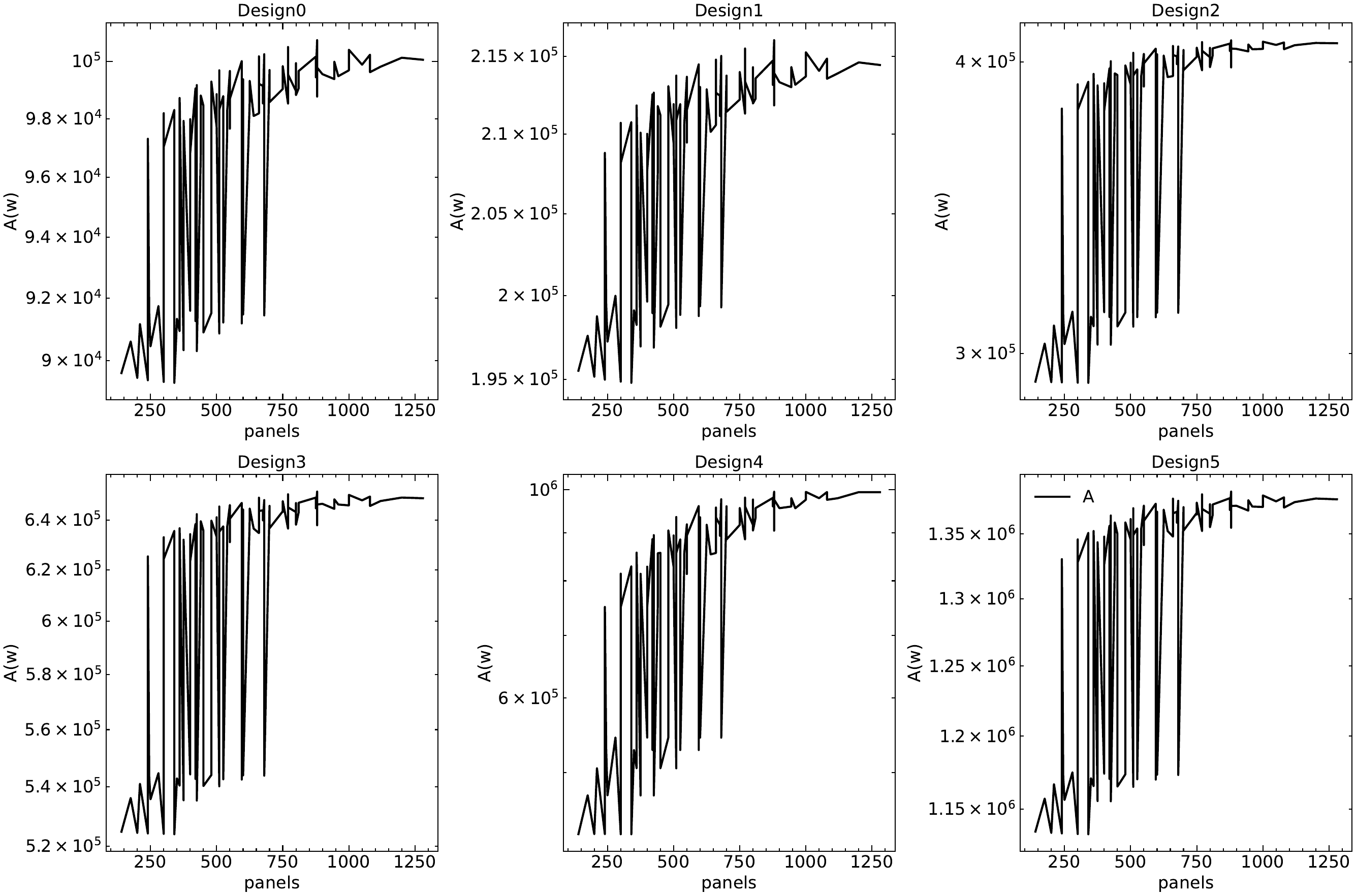}
    \caption{Mesh convergence for sample designs. The number of panel is shown on the x-axis and the $\mathbf{A}(\omega)$ is shown on the y-axis. All of these design converge for at least 750-1000 panels.}
    \label{fig:mesh_convergence}
\end{figure}
\begin{figure}[H]
    \centering \includegraphics[width = 0.7\linewidth]{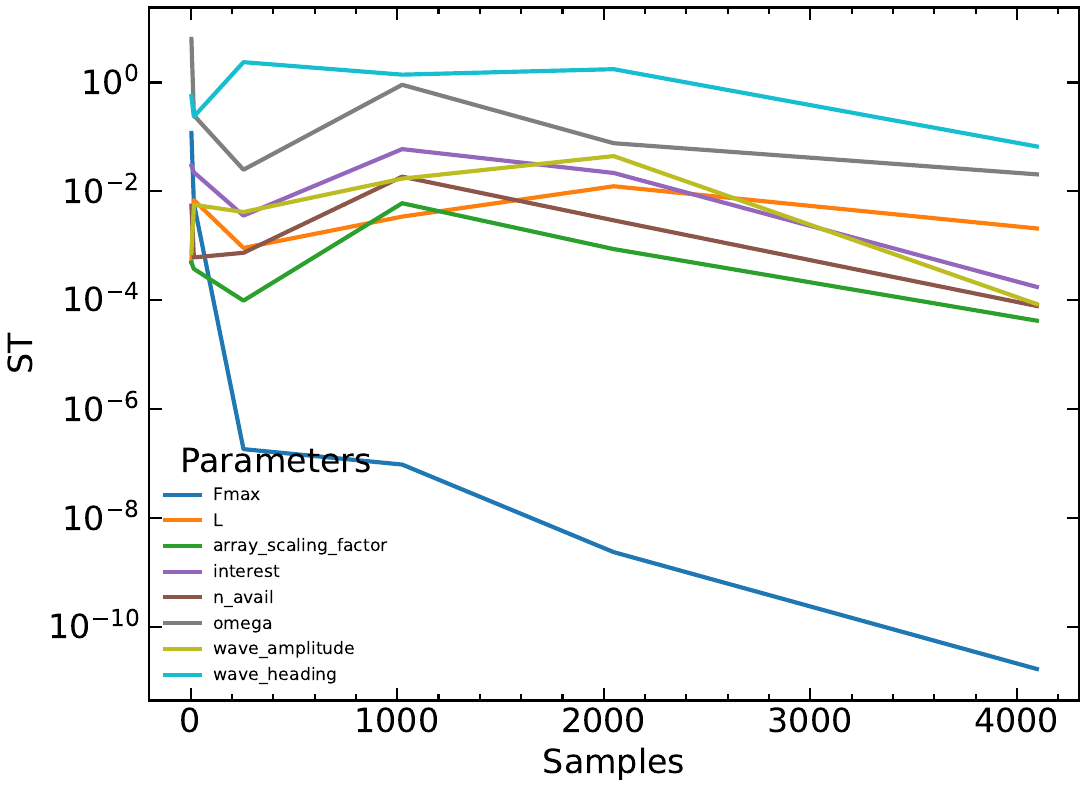}
    \caption{Sobol convergence}
    \label{fig:sobolConv}
\end{figure}

\section{Nominal Array Layouts}

\begin{figure}[H]
\centering
\includegraphics[width=0.7\linewidth]{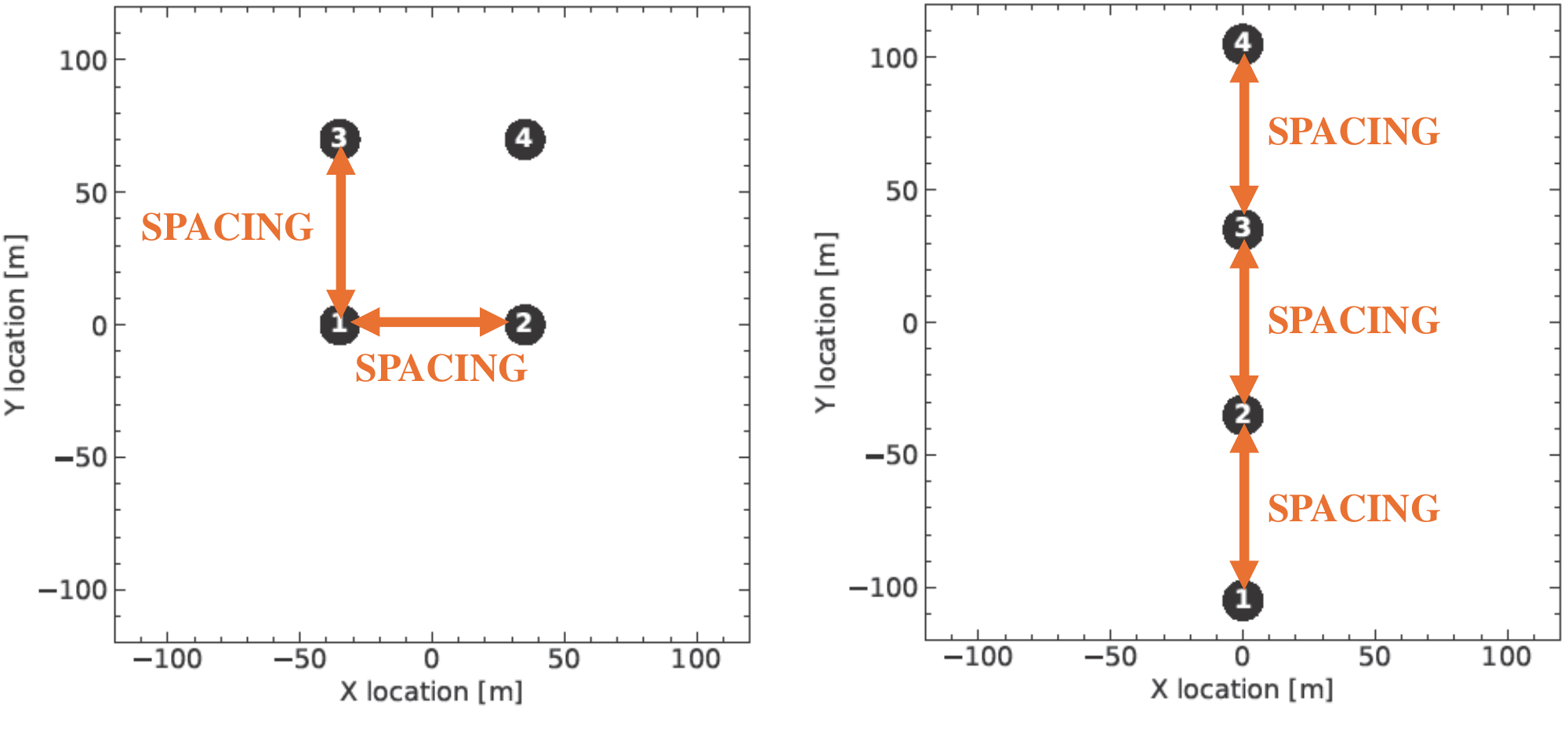}
\caption{Plots of nominal array layouts corresponding to the single objective optimization results in Table \ref{table:geom_damps_soo}. The left corresponds to ``grid" and the right is ``line".}
\label{fig:layouts}
\end{figure}

\begin{figure}[H]
    \centering
    \includegraphics[width = 0.5\linewidth]{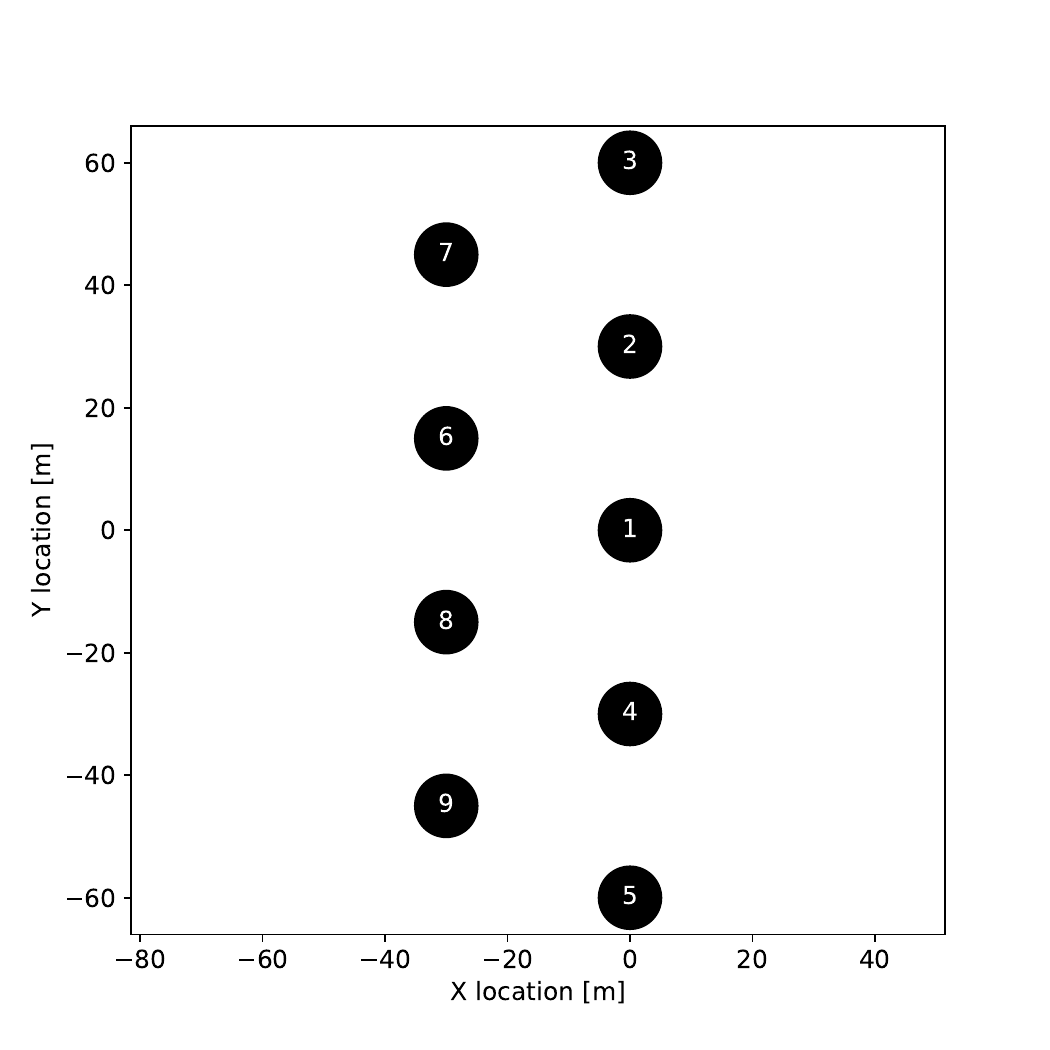}
    \caption{Layout of one cluster from Balitsky study, used for validation.}
    \label{fig:balit_layout}
\end{figure}

\begin{figure}[H]
    \centering
    \begin{subfigure}{0.45\textwidth}
        \centering
        \includegraphics[width=\textwidth]{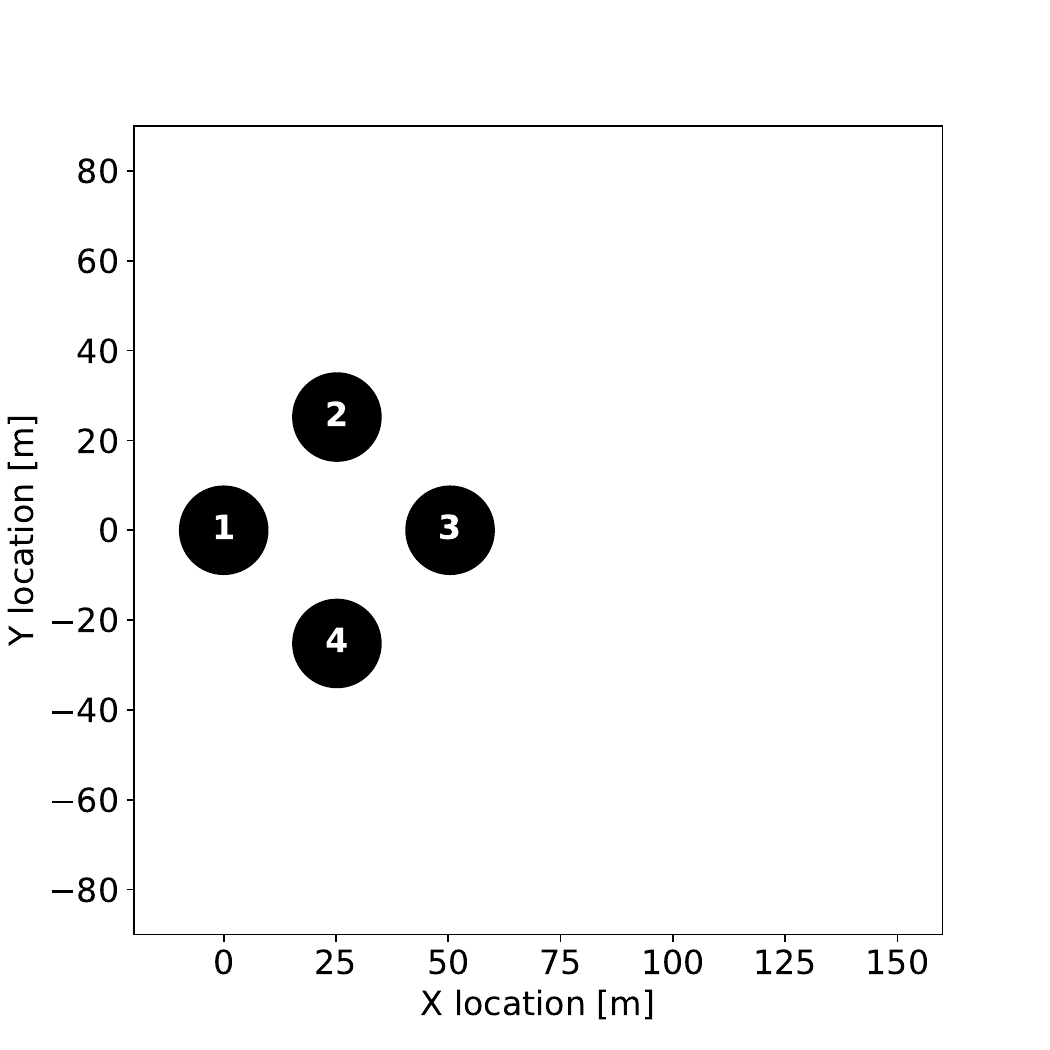}
        \caption{Diamond (tight)}
        \label{fig:diamondtight}
    \end{subfigure}
    \hfill
    \begin{subfigure}{0.45\textwidth}
        \centering
        \includegraphics[width=\textwidth]{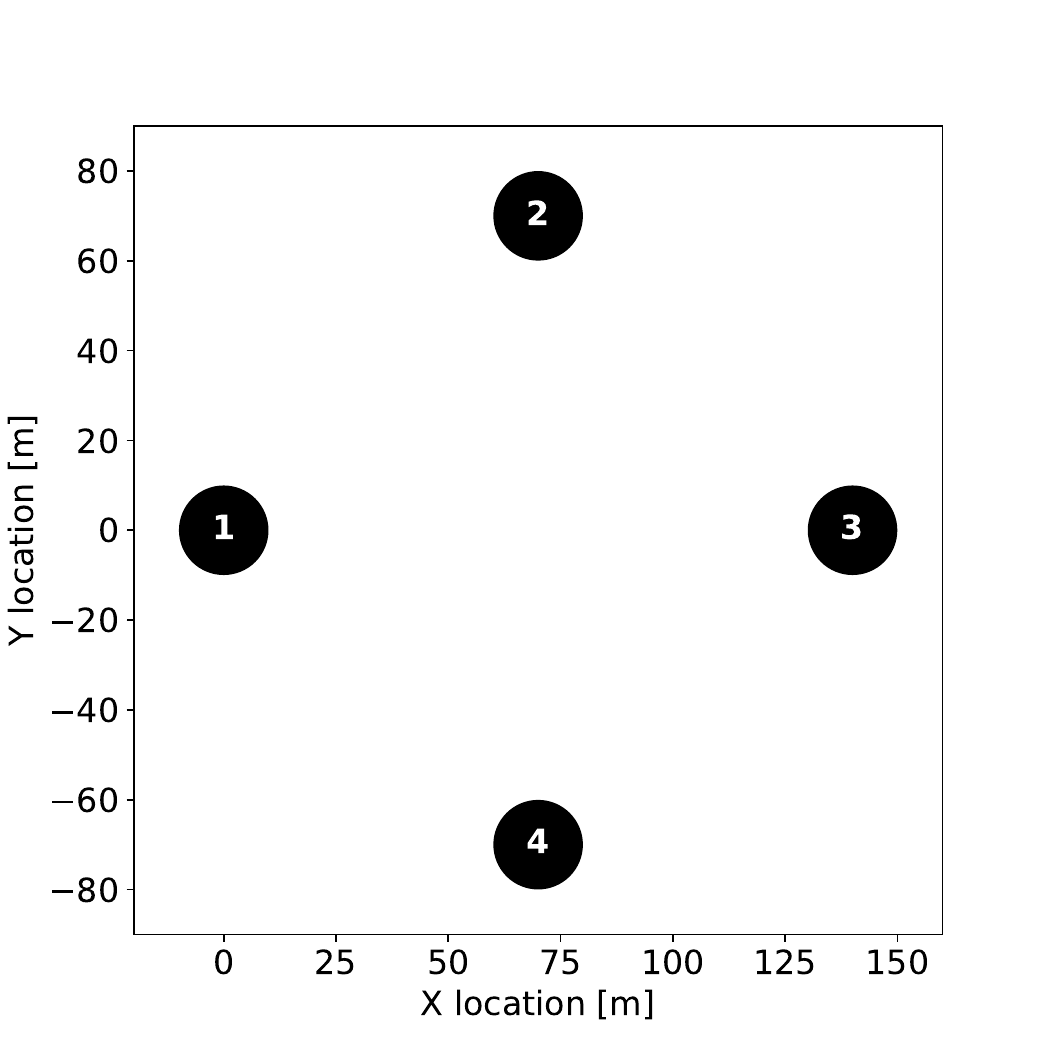}
        \caption{Diamond (spacious)}
        \label{fig:diamondspace}
    \end{subfigure}
    \caption{Diamond layouts mentioned in Table \ref{tbl:sat}.}
    \label{fig:diamond}
\end{figure}

\section{Unfiltered Pareto}
\begin{figure}[H]
    \centering
    \includegraphics[width = 0.7\linewidth]{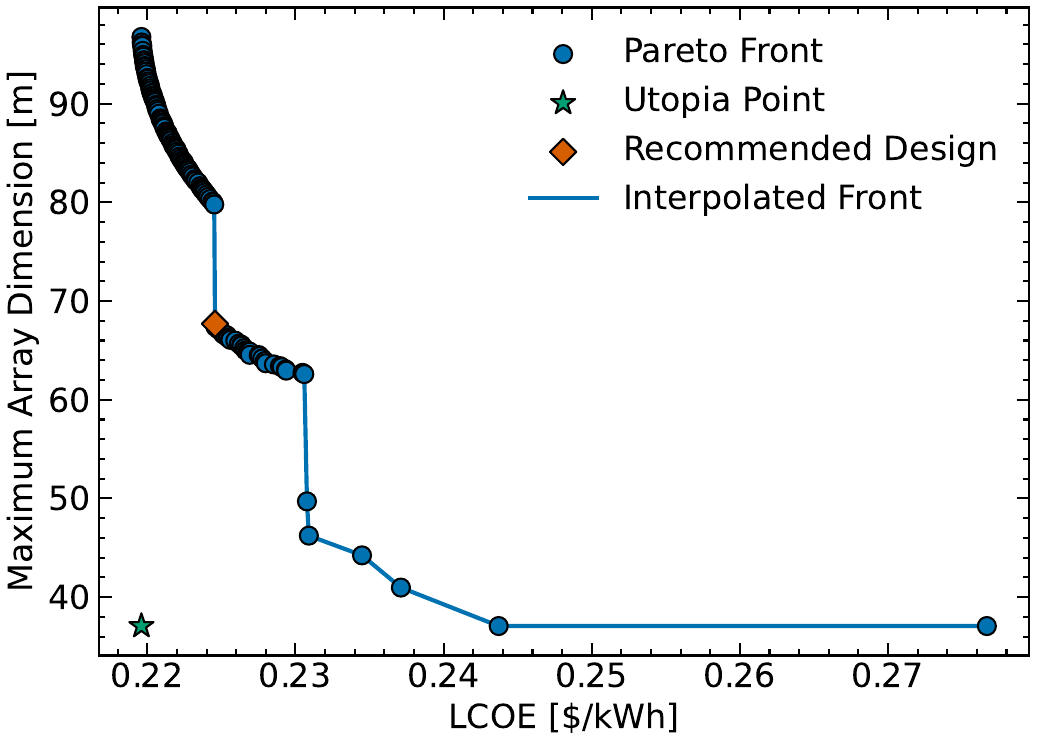}
    \caption{Plot of the unfiltered pareto front. The six points with the smallest Maximum Array Dimension are the points that got filtered.}
    \label{fig:pareto_full}
\end{figure}

\bibliographystyle{elsarticle-harv} 
 \bibliography{references}


\end{document}